\documentclass[prb,aps,twocolumn,superscriptaddress,floatfix]{revtex4}
\usepackage{graphicx}
\usepackage{amssymb}
\usepackage{amsmath}
\usepackage{xspace} 
\usepackage{color}

\begin{document}

\title{Exciton condensation due to electron-phonon interaction}
\author{N. V. Phan}
\affiliation{Institute of Physics, Vietnam Academy of Science and Technology, 10 Daotan, Badinh, Hanoi, Vietnam}
\author{Klaus W. Becker}
\affiliation{
Institut f{\"u}r Theoretische Physik, Technische Universit{\"a}t Dresden, D-01062 Dresden, Germany}
\author{Holger Fehske}
\affiliation{
Institut f{\"u}r Physik, Ernst-Moritz-Arndt-Universit{\"a}t Greifswald, D-17489 Greifswald, Germany}


\begin{abstract}
We show that the coupling to vibrational degrees of freedom can drive a semimetal excitonic-insulator quantum phase transition in an one-dimensional  two-band  $f$-$c$ electron system at zero temperature. The insulating state typifies an excitonic condensate accompanied by a finite lattice distortion. Using the projector-based renormalization method  we analyze the ground-state and spectral properties of the interacting electron-phonon model at half-filling. In particular we  calculate the momentum dependence of the excitonic order parameter function and determine the finite critical interaction strength for the metal-insulator transition to appear. The  electron spectral function reveals  the strong hybridization of  $f$- and $c$-electron states and the opening of a single-particle excitation gap. The phonon spectral function  indicates that the phonon mode involved in the transition softens (hardens) in the adiabatic (non-adiabatic and extreme anti-adiabatic) phonon frequency regime. 
\end{abstract}
\date{\today}
\maketitle

\section{Introduction}
Low-dimensional electron systems  are very susceptible to structural distortions driven by the electron-phonon interaction.\cite{TNYS91}  Probably the most famous one is the {\it Peierls instability} of one-dimensional  metals,\cite{Pe55} where the system spontaneously creates a periodic variation in the carrier density at any finite coupling by shifting the ions from their symmetric positions. For the half-filled band case  this so-called charge density wave (CDW) is commensurate with the lattice.  Since a static dimerisation of the lattice opens a gap at the Fermi surface the metal gives way to an insulator.  A full understanding of such a zero-temperature quantum phase transition requires accounting for both quantum lattice fluctuations and strong electronic correlations. For example, it has been found theoretically that quantum fluctuations of the lattice `protect'  the metallic state at weak electron-phonon couplings below a finite critical coupling strength.\cite{HF82} An intra-site Coulomb repulsion between electrons of opposite spin, on the other hand, tends to immobilize the carriers, but establishes a Mott insulating ground state with strong spin-density correlations instead of the CDW. To analyze the subtle interplay of electron-electron and electron-phonon interaction effects the one-band Holstein-Hubbard model turned out to be particularly rewarding to study.\cite{TC03}

If we have two electronic bands, however, forming a semimetal with only weak band overlap or a semiconductor with small band gap,  the Coulomb interaction between $f$-band (`hole') and $c$-band (`electron') particles  causes the formation of  (electron-hole) bound states. 
Then, at the semimetal-semiconductor transition, the ground state of the crystal may become unstable with respect to the spontaneous formation of excitons. That is,  an {\it excitonic instability}  appears,\cite{Mo61}  where the number of free carriers will vary discontinuously under an applied perturbation,  signalling a quantum phase transition. The new macroscopic phase-coherent quantum state 
can be regarded as an electron-hole pair condensate. Worth mentioning the excitonic state exhibits no `super' transport
properties;\cite{Ko66} rather it typifies an `excitonic insulator' (EI) which---under certain conditions---is accompanied 
by a CDW.\cite{HR68b} Such a density oscillation can, of course, trigger a lattice distortion which doubles the lattice period, just as for the Peierls state discussed above.

The challenging suggestion of electron-hole pair condensation into the EI phase at equilibrium  has been intensively studied  within the frameworks of purely electronic, effective-mass Mott-Wannier-type exciton\cite{BF06,CMCBDGBA07} 
and extended Falicov-Kimball models. \cite{Ba02b,Fa08,PBF10,ZIBF12}  
In doing, so the coupling to the phonons was  neglected.  Since the nondistorded semimetal ground state of a simple two-band model with electron and hole Fermi surfaces identical in size and shape is unstable with respect to  electron-hole attraction near the semimetal-semiconductor  transition, just as the normal Fermi surface of a metal is unstable to the formation of Cooper pairs,\cite{HR68b} one might ask whether the coupling of electrons and holes to the lattice degrees of freedom alone is sufficient to drive an EI instability. Addressing this question is the primary concern of this paper.  

Whether a CDW transition arising from the coupling between valence and conduction band electrons  is brought about by the electron-electron interaction or by the electron-phonon coupling has been debated for a number of materials in the recent past.  
For example, in spite of many experimental and theoretical studies, the origin of CDW  instability in transition-metal dichalcogenide 
1$T$-$\rm TiSe_2$ remains controversial: it could be the consequence of a novel  indirect Jahn-Teller effect,\cite{KMCC02} of  phonon softening,\cite{HZHCC01}   the formation of an EI condensate,\cite{CMCBDGBA07} or the combination\cite{MMAB12} 
of the latter both scenarios.\footnote{Recent experiments on 1$T$-$\rm TiSe_2$  point to a very unusual chiral property of the CDW;
see Ishioka {\it et. al.}   Phys. Rev. Lett. {\bf 105}, 176401 (2010).}  Also for the mixed-valent rare-earth chalcogenide $\rm TmSe_{0.45}Te_{0.55}$,\cite{NW90}  the lattice degrees of freedom seem to play an important role forming the EI state: very recent heat capacity measurements indicate that the excitons couple to phonons in the sense of exciton-polarons.\cite{WB13}

In this work, we study a one-dimensional two-band $f$-$c$ electron model with a coupling to the phonon degrees of freedom only and show that this interaction mediates a `hybridization' between $f$ and $c$ electrons.  From a theoretical point of view, we employ both a standard mean-field scheme and the projector-based renormalization method (PRM).\cite{BHS02} The PRM approach, described in Appendix~A, thereby includes fluctuation corrections. It enables the calculation of both ground state and spectral quantities for correlated many-particle systems,  and furthermore has the ability to find broken-symmetry  solutions of phase transitions beyond mean-field theory. We present our numerical results in Sec.~\ref{S:V}. Our conclusions can be found in Sec.~\ref{S:IV}.


\section{Model}
\label{S:II}
\begin{figure}[htb]
    \begin{center}
      \includegraphics[angle = -0, width = 0.4\textwidth]{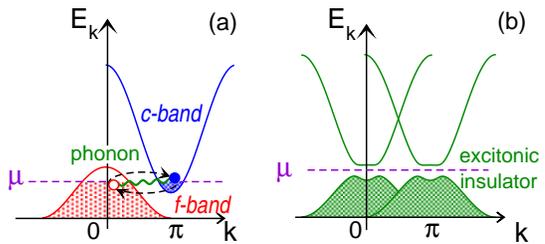}
    \end{center}
\caption{(Color online) Panel (a): Semimetallic $f$-$c$ electron band structure 
used in this work.  An $f$ valence-band hole and a $c$ conduction-band electron may form an `excitonic' bound state
owing to their interaction with the lattice degrees of freedom,  with a Brillouin zone boundary phonon involved 
for momentum conservation reasons. Note that the schematic band structure shown mimics the situation
in the EI material $\rm TmSe_{0.45}Te_{0.55}$, where the quasilocalized 4$f^{13}$ state has its maximum at the $\Gamma$ point, the 5$d$ strongly dispersive state has its minimum at the X point, and 
exciton formation is accompanied by a  $\Gamma$-X phonon.\cite{WB13}   Also in 1$T$-$\rm TiSe_2$ the valence-band top and the conduction-band minimum are located at different points in the Brillouin zone.\cite{ZF78}  
Panel (b): If  above a critical electron-phonon coupling strength  $f$-$c$ electron coherence is achieved  at sufficiently low temperatures even a new symmetry-broken ground state may appear, the so-called excitonic  insulator, which maybe accompanied  by a finite lattice distortion and a modulation of the charge density.\cite{HR68b}  }
\label{fig:D-G0}
\end{figure}

Let us consider the following coupled electron-phonon system \begin{eqnarray}\label{1}
\mathcal{H}&=&\sum_{\mathbf{k}}{\varepsilon}^f_{\mathbf{k}}f^\dagger_{\mathbf{k}}f^{}_{\mathbf{k}}
+\sum_{\mathbf{k}}{\varepsilon}^c_{\mathbf{k}}c^\dagger_{\mathbf{k}}c^{}_{\mathbf{k}}
+\omega_{0}\sum_{\mathbf{q}}b^\dagger_{\mathbf{q}}b^{}_{\mathbf{q}}\nonumber\\
&+& \frac{g}{\sqrt{N}}\sum_{\mathbf{kq}}\big(c^\dagger_{\mathbf{k+q}}f^{}_{\mathbf{k}}
( b^{\dagger}_{\mathbf{-q}}+ b_{\mathbf q} ) + \textrm{H.c} \big) \,,
\end{eqnarray}
which contains two types of spinless electrons $(c,f)$ carrying momentum $\bf k$ and 
dispersionless phonons ($b$) (see Fig.~\ref{fig:D-G0}). 
Here, the electronic excitation energies  are given by
\begin{equation}
\label{2}
{\varepsilon}^{f,c}_{\mathbf{k}}=\varepsilon_{}^{f,c}
-t^{f,c}_{}\gamma_{\mathbf k}-\mu \, ,
\end{equation}
whereas $\omega_0$ is the dispersionless phonon energy. In Eq.~\eqref{2},  
$\varepsilon^{f,c}$ represents the local part of the respective electronic excitations, and the   
term $-t^{f,c} \gamma_{\mathbf k}$, with $ \gamma_{\mathbf k}=2\cos k$, accounts for a  
nearest-neighbor hopping  in a one-dimensional lattice. 
Hereafter all energies are given in units of $t^c=1$. We furthermore note that the electronic energies are 
measured from the chemical potential $\mu$; where the numerical 
results presented in Sec.~\ref{S:V} contain an additional energy shift by fixing  $\varepsilon^c=0$.
The last term in Eq.~\eqref{1} describes a local electron-phonon interaction (with coupling constant $g$),
written in ${\bf k}$-space, between local $f$-$c$ particle-hole excitations and lattice displacements.
Apparently it represents an effective `exciton'-phonon interaction.  In Eq.~\eqref{1} we have introduced Fourier  
transformed quantities $f^\dag_{\bf k} =(1/\sqrt N) \sum_i f^\dag_i e^{i\bf k {\bf R}_i}$,  $c^\dag_{\bf k} =(1/\sqrt N) \sum_i c^\dag_i e^{i\bf k {\bf R}_i}$, and $b^\dag_{\bf q} =(1/\sqrt N) \sum_i b^\dag_i e^{i\bf q {\bf R}_i}$, where
$f^\dag_i, c^\dag_i$ and $b^\dag_i$ are the local quantities. 
 $N$ counts the number of lattice sites~$i$. 

 In what follows, we consider a half-filled band, i.e., 
 \begin{equation}
 \label{3}
n=\langle n^f_i\rangle+\langle n^c_i\rangle =1\,,
\end{equation}
where $n^f_i=({1}/{N})\sum_{\bf k} f^\dagger_{\bf k}f^{}_{\bf k}$, $n^c_i
=({1}/{N})\sum_{\bf k} c^\dagger_{\bf k}c^{}_{\bf k}$. The chemical potential $\mu$ has to be adjusted 
in such a way that Eq.~\eqref{3} is satisfied. 
Without loss of generality, in what follows, the $c$ electrons will be considered 
as `light' while the $f$ electrons (respectively holes) are `heavy', i.e. $|t^f|<1$. 
For negative $t^f$, and coinciding energies of $c$ and $f$ electrons, one is led to a 
picture of indirect $c-f$ hopping (cf. Fig.~\ref{fig:D-G0}), which suggests 
a possible condensation of bound $c$-$f$ electron-hole pairs with finite momentum:
\begin{equation}
\label{3}
d_{\bf k}= \langle c^\dag_{\bf k + \bf Q} f_{\bf k}^{} \rangle \neq 0 \, ,
\end{equation}
where $\bf Q= \pi$ in one dimension.  Allowing broken symmetry solutions for non-vanishing $d_{\bf k}$, 
small infinitesimal fields must be included in model \eqref{1}. We write
\begin{align}
\label{4}
\mathcal{H}&=
\sum_{\mathbf{k}}{\varepsilon}^f_{\mathbf{k}}f^\dagger_{\mathbf{k}}f^{}_{\mathbf{k}}
+\sum_{\mathbf{k}}{\varepsilon}^c_{\mathbf{k}}c^\dagger_{\mathbf{k}}c^{}_{\mathbf{k}}
+\omega_{0}\sum_{\mathbf{q}}b^\dagger_{\mathbf{q}}b^{}_{\mathbf{q}}\nonumber\\
&+ \Delta_0 \sum_{\bf k} \big( c^\dag_{\bf k +\bf Q} f_{\bf k} + f^\dag_{\bf k} c_{\bf k + \bf Q} \big)
+ \sqrt N h_0 \big( b_{-\bf Q}^\dag + b_{ \bf Q} \big) \nonumber \\ 
&+ \frac{g}{\sqrt{N}}\sum_{\mathbf{kq}}\big(c^\dagger_{\mathbf{k+q}}f^{}_{\mathbf{k}}
( b^{\dagger}_{\mathbf{-q}}+ b_{\mathbf q} ) + \textrm{H.c} \big) 
 \,, 
\end{align}
where $\Delta_0=0^+$ and $h_0=0^+$. 
It is easily realized that the fields $h_0$ and $\Delta_0$ are mutually dependent. 
Moreover, since $b_{\bf Q} = b_{-\bf Q}$ the field contribution
$\sqrt N h_0 \big( b_{-\bf Q}^\dag + b_{ \bf Q} \big)$ can be replaced by 
$\sqrt N h_0 \big( b_{-\bf Q}^\dag + b_{ -\bf Q} \big)$. Therefore a finite 
lattice displacement $\propto  \langle b_{-\bf Q}^\dag + b_{ -\bf Q}\rangle$ would 
gives rise to the formation of a charge density wave connected to a doubling of the lattice unit cell.  \\


\section{Mean-field theory}
\label{S:III}

To solve model \eqref{4} in mean-field approximation it is advantageous first to introduce 
fluctuation operators $\delta \mathcal A= \mathcal A - \langle \mathcal A\rangle$
in the electron(exciton)-phonon interaction. Using 
\begin{align}
\label{5}
&\delta(c^\dagger_{\mathbf{k+q}}f^{}_{\mathbf{k}})\delta(b^\dagger_{\mathbf{-q}} +  b_{\bf q})=
c^\dagger_{\mathbf{k+q}}f^{}_{\mathbf{k}} (b^\dagger_{\mathbf{-q}} +  b_{\bf q}) \nonumber \\
&- \big[\langle c^\dagger_{\mathbf{k+q}}f^{}_{\mathbf{k}}\rangle (b^\dagger_{\mathbf{-q}} +  b_{\bf q})
+c^\dagger_{\mathbf{k+q}}f^{}_{\mathbf{k}} \, \langle b^\dagger_{\mathbf{-q}} +  b_{\bf q} \rangle \big]
 \delta_{\bf q, \bf Q}
\nonumber \\
&+\langle c^\dagger_{\mathbf{k+q}}f^{}_{\mathbf{k}}\rangle \,  \langle b^\dagger_{\mathbf{-q}} +  b_{\bf q}\rangle
\delta_{\bf q, \bf Q}\,,
\end{align}
 the Hamiltonian $\mathcal H$ is best rewritten as  
 \begin{equation}
\label{6}
 \mathcal H= \mathcal H_0 + \mathcal H_1 
\end{equation}
with
 \begin{align}
 \label{7}
\mathcal{H}_0&=\sum_{\mathbf{k}}{\varepsilon}^f_{\mathbf{k}}f^\dagger_{\mathbf{k}}f^{}_{\mathbf{k}}
+\sum_{\mathbf{k}} {\varepsilon}^c_{\mathbf{k}}c^\dagger_{\mathbf{k}}c^{}_{\mathbf{k}}
+\omega_{0}\sum_{\mathbf{q}}b^\dagger_{\mathbf{q}}b^{}_{\mathbf{q}}\\
&+  \Delta \sum_{\bf k} \big( c^\dag_{\bf k +\bf Q} f_{\bf k} + f^\dag_{\bf k} c_{\bf k + \bf Q} \big)
+ \sqrt N h \big( b_{-\bf Q}^\dag + b_{- \bf Q} \big)\,, \nonumber \\ 
\mathcal{H}_1& =
\frac{g}{\sqrt{N}}\sum_{\mathbf{kq}} \Big[\delta(c^\dagger_{\mathbf{k+q}}f^{}_{\mathbf{k}})
\delta( b^{\dagger}_{\mathbf{-q}}+ b_{\mathbf q} ) + \textrm{H.c} \Big]  \, .
\end{align}
 Here the fields have acquired additional shifts, which will act as order parameters in the 
 following:
 \begin{eqnarray}
 \label{8}
 && \Delta= \Delta_0 + \frac{g}{\sqrt N} \langle b_{-\bf Q} + b^\dag_{-\bf Q} \rangle \,, \\
&& h  = h_0 + \frac{g}{N} \sum_{\bf k} \langle c^\dagger_{\mathbf{k+Q}}f^{}_{\mathbf{k}}
+ f_{\mathbf{k}}^\dag c_{\mathbf{k+Q}} \rangle \, ,
 \label{8a}
 \end{eqnarray}
where the infinitesimal $\Delta_0= 0^+$ and $h_0=0^+$ 
can be neglected for finite expectation values on the right hand sides. \\

Finally, we eliminate in Eq.~\eqref{7}  the term $\propto \big( b_{-\bf Q}^\dag + b_{- \bf Q} \big)$ by defining  
new phonon operators 
\begin{equation}
\label{npo}
B_{\bf q}^\dag = b_{\bf q}^\dag +   \sqrt N ({h}/{\omega_0}) \delta_{\bf q, \bf Q}\,,
\end{equation} where the definition is independent 
of the sign of $\bf Q$. 
$\mathcal H_0$ and $\mathcal H_1$ then become
\begin{align}
 \label{9}
\mathcal{H}_{0}&=\sum_{\mathbf{k}}{\varepsilon}^f_{\mathbf{k}}f^\dagger_{\mathbf{k}}f^{}_{\mathbf{k}}
+\sum_{\mathbf{k}}{\varepsilon}^c_{\mathbf{k}}c^\dagger_{\mathbf{k}}c^{}_{\mathbf{k}}
+ \omega_{0}\sum_{\mathbf{q}} B^\dagger_{\mathbf{q}} B^{}_{\mathbf{q}}\\
&+  \Delta \sum_{\bf k} \big( c^\dag_{\bf k +\bf Q} f_{\bf k} + f^\dag_{\bf k} c_{\bf k + \bf Q} \big)
+{\rm const.}\,, \nonumber \\
\label{9a}
 \mathcal H_{1} &=
\frac{g}{\sqrt{N}}\sum_\mathbf{kq} \Big[ \delta(c^\dagger_{\mathbf{k+q}}f^{}_{\mathbf{k}}) \,
\delta( B^{\dagger}_{\mathbf{-q}}+ B_{\mathbf q} ) + \textrm{H.c} \Big]  \, .
\end{align}
In $\mathcal H_{1}$ we have used $\delta B^\dag_{-\bf q}= \delta b^\dag_{-\bf q}$ and 
$\delta B_{\bf q}= \delta b_{\bf q}$.   \\

Note that the Hamiltonian $\mathcal H= \mathcal H_0 + \mathcal H_1$, with $\mathcal H_0$ and 
$\mathcal H_1$ given by Eqs.~\eqref{9} and \eqref{9a}, is still exact.
The Hamiltonian in mean-field approximation is obtained by completely neglecting the fluctuation part $\mathcal H_1$. 
Thus the mean-field Hamiltonian reads 
 \begin{align}
 \label{9b}
\mathcal H_{MF} &= \sum_{\mathbf{k}}{\varepsilon}^f_{\mathbf{k}}f^\dagger_{\mathbf{k}}f^{}_{\mathbf{k}}
+\sum_{\mathbf{k}}{\varepsilon}^c_{\mathbf{k}}c^\dagger_{\mathbf{k}}c^{}_{\mathbf{k}}
+ \omega_{0}\sum_{\mathbf{q}} B^\dagger_{\mathbf{q}} B^{}_{\mathbf{q}} \nonumber \\
&+  \Delta \sum_{\bf k} \big( c^\dag_{\bf k +\bf Q} f_{\bf k} + f^\dag_{\bf k} c_{\bf k + \bf Q} \big) \, ,
 \end{align} 
 where the constant from Eq.~\eqref{9} will be suppressed. 
The electronic part of ${\mathcal H}_{MF}$ is diagonalized by use of a 
Bogoliubov transformation. Then $\mathcal H_{MF}$ is rewritten as
\begin{eqnarray}
\label{9c}
{\mathcal{H}}_{MF}&=&\sum_{\mathbf{k}}E^{(1)}_{\mathbf{k}}{C}^{\dagger}_{1, \mathbf{k}}{C}_{1, \mathbf{k}}
+\sum_{\mathbf{k}}E^{(2)}_{\mathbf{k}}{C}^{\dagger}_{2, \mathbf{k}}{C}_{2, \mathbf{k}} \nonumber \\
&+& {\omega}_{0} \sum_{\bf q}  {B}^\dag_{\bf q} {B}_{\bf q} \, ,
\end{eqnarray}
where the electronic quasiparticle energies and quasiparticles operators read
\begin{eqnarray}
\label{9d}
E^{(1,2)}_{\mathbf k}&=&\frac{{\varepsilon}^c_\mathbf{k+Q}+{\varepsilon}^f_{\mathbf k}}{2}
\mp
\frac{\textrm{sgn}({\varepsilon}^f_{\mathbf k}-{\varepsilon}^c_\mathbf{k+Q})}
{2}W_{\mathbf k}\,,
\end{eqnarray}
and 
\begin{eqnarray}
\label{9e}
{C}^\dagger_{1, \mathbf{k}}&=&\xi^{}_{\mathbf{k}}c^\dagger_{\mathbf{k+Q}}+ \eta^{}_{\mathbf{k}}f^\dagger_{\mathbf{k}}\, , \\
{C}^\dagger_{2, \mathbf{k}}&=-&\eta^{}_{\mathbf{k}}c^\dagger_{\mathbf{k+Q}}+\xi^{}_{\mathbf{k}}f^\dagger_{\mathbf{k}} \, .
\end{eqnarray}
Here the prefactors are given by 
\begin{eqnarray}
\label{9f}
\xi^{2}_{\mathbf{k}}&=&\frac{1}{2}\left[1+\textrm{sgn}({\varepsilon}^f_{\mathbf k}-{\varepsilon}^c_\mathbf{k+Q})
\frac{{\varepsilon}^f_{\mathbf k}-{\varepsilon}^c_\mathbf{k+Q}}{W_{\mathbf{k}}}\right]\,, \\
\eta^{2}_{\mathbf{k}}&=&\frac{1}{2}\left[1-\textrm{sgn}({\varepsilon}^f_{\mathbf k}-{\varepsilon}^c_\mathbf{k+Q})
\frac{{\varepsilon}^f_{\mathbf k}-{\varepsilon}^c_\mathbf{k+Q}}{W_{\mathbf{k}}}\right] \, ,
\label{9fa}
\end{eqnarray}
with
\begin{eqnarray}
\label{9g}
W_{\mathbf k}=\sqrt{({\varepsilon}^c_{\mathbf{k+Q}}-{\varepsilon}^f_{\mathbf k})^2+4|{\Delta}|^2}\,.
\end{eqnarray}
The quadratic form of Eq.~\eqref{9c} allows to compute all expectation values formed with 
${\mathcal H}_{MF}$.  From Eqs.~\eqref{8}, \eqref{8a}, and \eqref{npo} one easily obtains the following implicit equation 
for the order parameters $\Delta= -(2g/\omega_0) h$,
\begin{eqnarray}
\label{9h}
1= \frac{4g^2}{\omega_0} \frac{1}{N} \sum_{\bf k} \textrm{sgn} (\varepsilon^f_{\bf k} - \varepsilon^c_{\bf k + \bf Q})
\frac{f^F(E_{\bf k}^{(1)})- f^F(E_{\bf k}^{(2)})}{W_{\bf k}} \, . \nonumber \\
&&
\end{eqnarray}
Here $f^F(E_{\bf k}^{(1,2)})$ are Fermi functions, which---working at zero temperature in what follows---reduce 
to the corresponding $\Theta$-functions.
Note that Eq.~\eqref{9h} represents a BCS-like equation for $\Delta$. A non-zero $\Delta$ accounts for a
exciton  condensation phase as was explained above.  In Figs.~\ref{fig:D-T} and~\ref{fig:pd} below, it will be shown that such a phase 
occurs for a sufficiently large coupling constant $g > g_c^{MF}(\omega_0)$. We would like to point out here already, that the critical coupling constant
 $g_c^{MF}$ is generally smaller than the corresponding $g_c^{PRM}$, obtained below by including fluctuation processes. 
 
Let us also consider the one-particle spectral function $A^{(c,f)}_{\bf k}( \omega)$ for $c$ and $f$ electrons. For  
$c$ electrons it is defined by 
 \begin{eqnarray}
\label{9i}
 A^c_{\bf k}(\omega) &=& \frac{1}{2\pi} \int_{-\infty}^\infty \langle [c_{\bf k \sigma}(t), c^\dag_{\bf k \sigma}]_+\rangle
 e^{i\omega t} dt \, , 
\end{eqnarray}
where the expectation values is formed with $\mathcal H_{MF}$. For $A^{c}(\bf k, \omega)$
 and the corresponding equation for the $f$ electrons one finds  
 \begin{eqnarray}
\label{9j}
 A^c_{\bf k}(\omega) &=& \xi_{\bf k- \bf Q}^2 \delta(\omega -E_{\bf k -\bf Q}^{(1)}) +
 \eta_{\bf k- \bf Q}^2 \delta(\omega -E_{\bf k -\bf Q}^{(2)}) \nonumber \\
 && \\
  \label{9k}
  A^f_{\bf k}(\omega) &=& \eta_{\bf k}^2 \delta(\omega -E_{\bf k -\bf Q}^{(1)}) +
 \xi_{\bf k}^2 \delta(\omega -E_{\bf k -\bf Q}^{(2)})  \, .
 \end{eqnarray}
Thus both spectral functions are built up by two coherent excitations
with energies $E_{\bf k}^{(1)}$ and  $E_{\bf k}^{(2)}$.
Finally, the phonon spectral function
 \begin{eqnarray}
\label{9l}
C_{\bf q}(\omega) &=& \frac{1}{2\pi \omega}\int_{-\infty}^\infty \langle [b_{\bf q}(t), b^\dag_{\bf q}] \rangle
 e^{i\omega t} dt  
\end{eqnarray}
is given by
 \begin{eqnarray}
\label{9m}
C_{\bf q}(\omega) &=& \frac{\delta(\omega -\omega_0)}{\omega_0} \,,
\end{eqnarray}
which shows a $\bf q$-independent excitation at $\omega= \omega_0$. Note that in contrast to the 
electronic excitations in Eqs.~\eqref{9j} and \eqref{9k}, the phonon frequency $\omega_0$ in not changed  
in mean-field approximation.


\section{Fluctuation corrections beyond  mean-field theory}
\label{S:IV}

In the mean-field treatment above fluctuation processes from the interaction $\mathcal H_1$ have completely been 
left out. In the following, we therefore apply the PRM\cite{BHS02} 
to evaluate the order parameters, the one-particle spectral functions $A^{(c,f)}_{\bf k}( \omega)$ and the phonon spectral function 
$C_{\bf q}(\omega)$ for the case that $\mathcal H_1$ is included. To avoid technical details, the explicit application is shifted  
to appendix \ref{A}.  The general concept of the PRM is as follows: The presence of the interaction $\mathcal H_1$
prevents a straightforward solution of the Hamiltonian $\mathcal H= \mathcal H_0 + \mathcal H_1$. 
For that reason the Hamiltonian is transformed into a diagonal (or at least quasi-diagonal) form by applying
a sequence of small unitary transformations to $\mathcal H$. Denoting for a moment the generator   
of the whole sequence by $X= -X^\dag$, in appendix \ref{A} it is shown that one can arrive at an effective Hamiltonian 
$\tilde{\mathcal H}= e^X \mathcal He^{-X}$, which has  
the same operator structure as Hamiltonian $\mathcal H_0$  in Eq.~\eqref{9}, 
 \begin{eqnarray}
\label{10a}
\tilde{\mathcal{H}}&=&\sum_{\mathbf{k}}\tilde{\varepsilon}^f_{\mathbf{k}}f^\dagger_{\mathbf{k}}f^{}_{\mathbf{k}}
+\sum_{\mathbf{k}}\tilde{\varepsilon}^c_{\mathbf{k}}c^\dagger_{\mathbf{k}}c^{}_{\mathbf{k}}
+ \sum_{\mathbf{q}} \tilde{\omega}_{\bf q}B^\dagger_{\mathbf{q}} B^{}_{\mathbf{q}} \nonumber \\
&+&   \tilde{\Delta}\sum_{\bf k} \big( c^\dag_{\bf k +\bf Q} f_{\bf k} +  f^\dag_{\bf k} c_{\bf k + \bf Q} \big) \, .
\end{eqnarray}
Here, $\tilde{\varepsilon}^f_{\mathbf{k}}$, $\tilde{\varepsilon}^c_{\mathbf{k}}$, $\tilde{\omega}_{\bf q}$, 
and $\tilde{\Delta}$ are renormalized parameters, which have to be determined self-consistently by taking into account contributions to infinite order in the interaction $\mathcal H_1$. Also, the phonon frequency $\tilde{\omega}_{\bf q}$ 
has acquired a $\bf q$-dependence. Note that the PRM ensures a well-controlled 
disentanglement of higher order interaction terms which enter in the elimination procedure. 

The PRM also allows to evaluate expectation values $\langle  \mathcal A \rangle$, formed with the full Hamiltonian 
$\mathcal H$. 
Thereby, one uses the property of unitary invariance of operator expressions 
under a trace. Employing the same unitary transformation to $\mathcal A$ as before for 
the Hamiltonian, one finds $\langle {\mathcal A} \rangle =
 \langle \tilde{\mathcal A} \rangle_{\tilde{\mathcal H}}$,  where the expectation value is formed with 
 $\tilde{\mathcal H}$ and $\tilde{\mathcal A} = e^X \mathcal A e^{-X}$.  Note that Hamiltonian $\tilde{\mathcal H}$ from Eq.~\eqref{10a}  can be transformed 
into a diagonal form by use of a Bogoliubov transformation in analogy to the transformation from Eq.~\eqref{9b} to Eq.~\eqref{9c}. Therefore, any expectation value, 
formed with  $\tilde{\mathcal H}$,
can be evaluated exactly. 

As an example, let us consider the spectral function $A^c_{\bf k}( \omega)$ from the former expression \eqref{9i}, where
the expectation value should however be formed with the full Hamiltonian ${\mathcal H}$ (and not with $\mathcal H_{MF})$. 
 Applying the  unitary invariance of operator expressions under a trace,  $A^c_{\bf k}( \omega)$ is rewritten as   
 \begin{eqnarray}
\label{10b}
A^c_{\bf k}( \omega) &=&  \frac{1}{2\pi} \int_{-\infty}^\infty 
\langle [\tilde{c}_{\bf k \sigma}(t), \tilde{c}^\dag_{\bf k \sigma}]_+\rangle_{\tilde{\mathcal H}}  e^{i\omega t} dt  \, ,
\end{eqnarray}
where the expectation value is now formed with  $\tilde{\mathcal H}$ instead of with $\mathcal H$.
Correspondingly $\tilde{c}^\dag_{\bf k \sigma}, \tilde{c}_{\bf k \sigma}$ are the transformed electron operators, 
$\tilde{c}^{(\dag)}_{\bf k \sigma}= e^X {c}^{(\dag)}_{\bf k \sigma} e^{-X}$, and the time-dependence is 
governed by $\tilde{\mathcal H}$ as well.


\section{Numerical results}
\label{S:V}
\subsection{Ground-state properties}
\label{S:V.A}
\begin{figure}[t]
    \begin{center}
     \includegraphics[angle = -90, width = 0.236\textwidth]{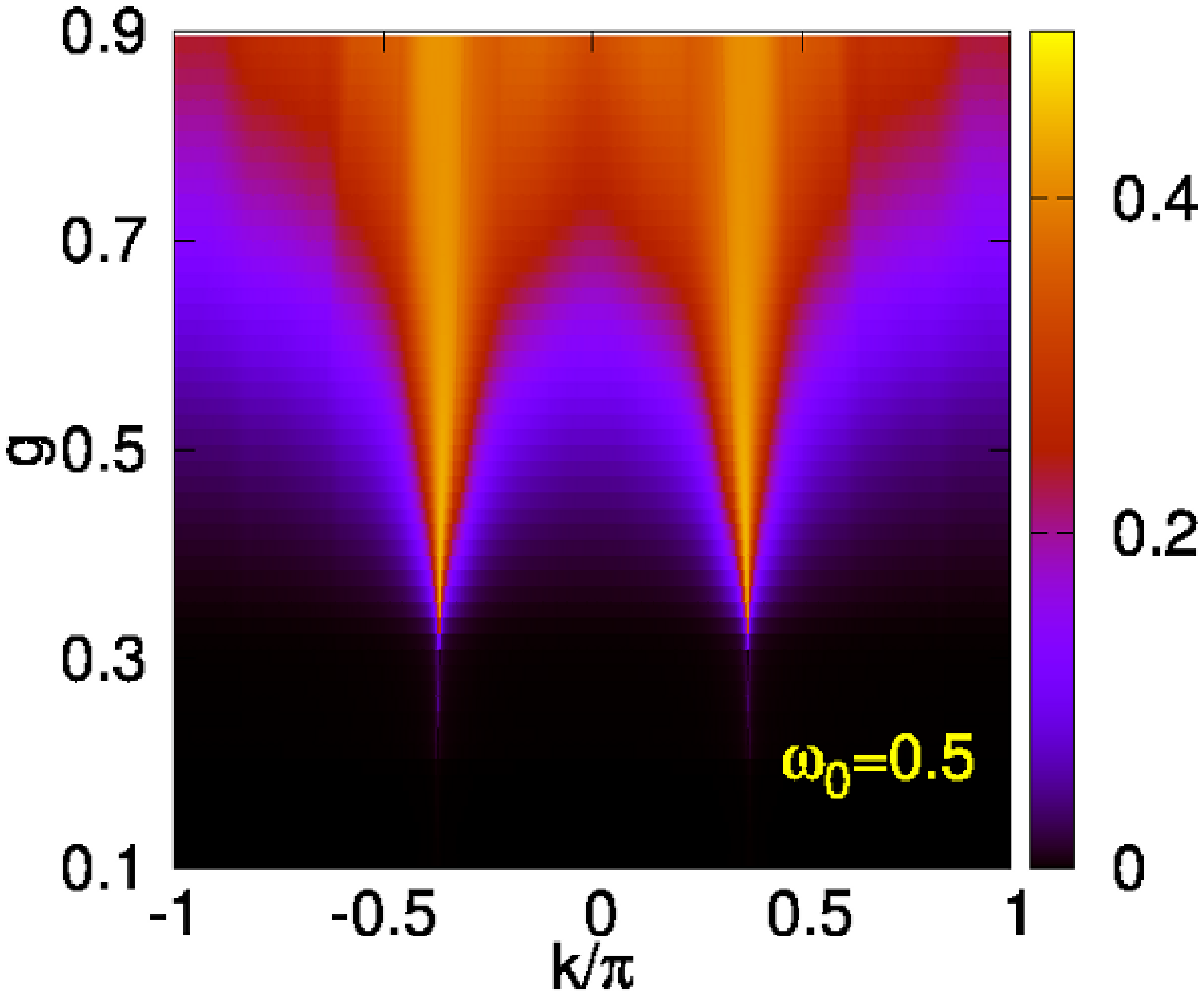}
     \includegraphics[angle = -90, width = 0.236\textwidth]{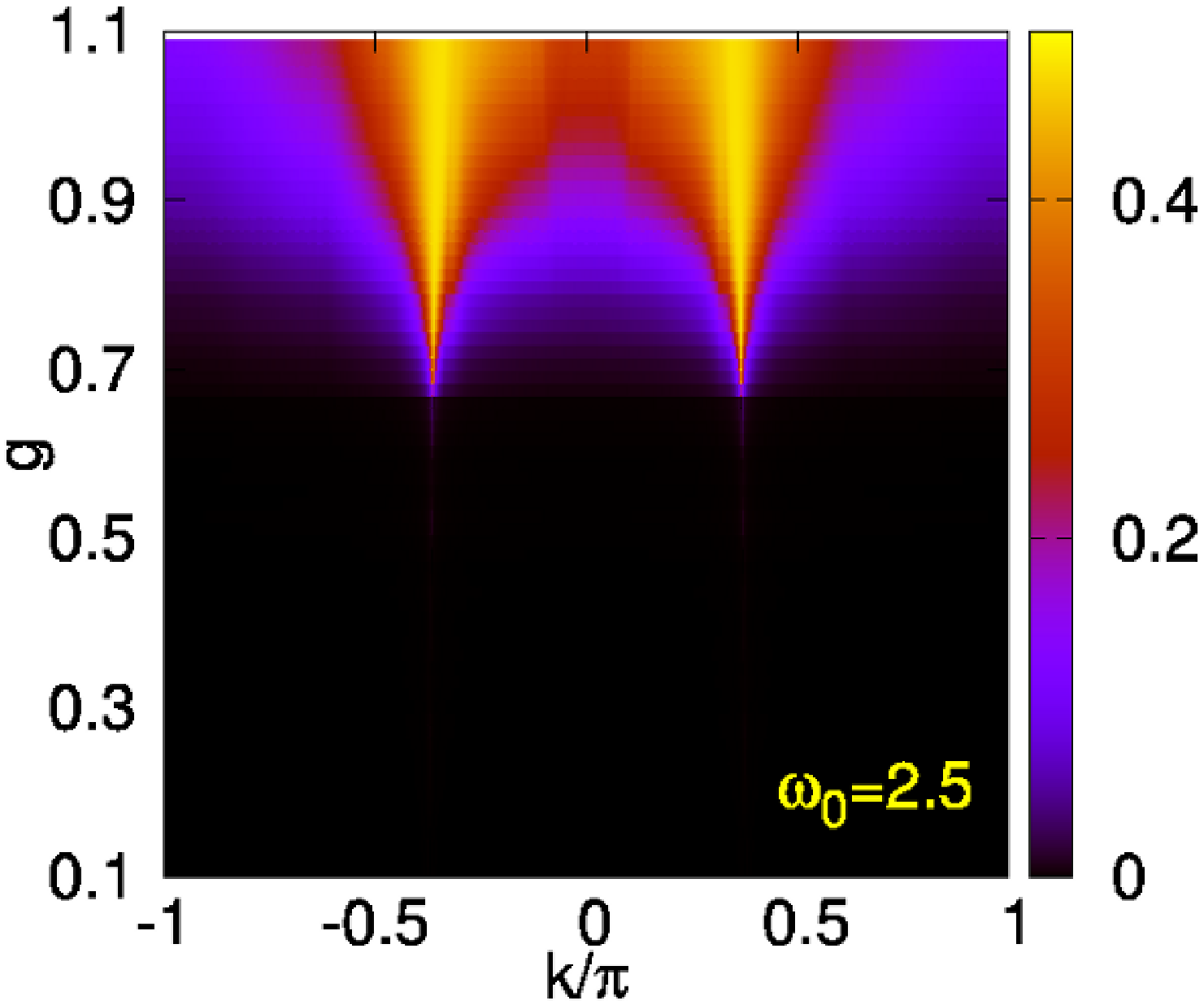}
    \end{center}
\caption{(Color online) Magnitude  of the EI order parameter function $d_k=\langle c^\dagger_{k+Q}f^{}_{k}\rangle$ with $Q=\pi$ (cf. color bar)
depending on momentum $k$ (on the $x$ axis) and  the electron-phonon coupling strength $g$ (on the $y$ axis) in  the adiabatic  
($\omega_0=0.5$, left-hand panel) and non-adiabatic 
($\omega_0=2.5$, right-hand panel) phonon frequency regime.}
\label{fig:Dk1}
\end{figure}

We start with a discussion of the EI order parameter $\Delta$ and the corresponding lattice displacement $x_Q$ in the 
ground state of the fully renormalized two-band model~\eqref{25} in one dimension. Figure~\ref{fig:Dk1} at first displays the profile of the excitonic 
expectation value $d_{k}$, in dependence on the electron-phonon coupling  $g$, for two characteristic phonon frequencies 
$\omega_0$ describing an adiabatic ($\omega_0<1$)  non-adiabatic ($\omega_0 > 1$) situation. This quantity designates the range in momentum space  
where $c$ electrons and $f$ holes are perceptibly involved in the electron-hole pair formation and exciton condensation process.
Obviously $d_{k}$ vanishes (within numerical accuracy) for all ${k}$ below a critical coupling strength [$g_c\simeq 0.28$ (0.6) at  $\omega_0=0.5$ (2.5)]. 
At and just above the critical coupling $d_{k}$ is solely finite at and near the Fermi momentum $k_{\rm F}$, respectively, indicating 
a BCS-type electron-hole pairing instability. A further increasing electron-phonon coupling implicates more and more electron and 
holes states  in the pairing process up to the point where Fermi surface (nesting) effects are ineffectual. Thus we expect that local,  
tightly bound excitons will form in the strong interaction limit and, as a consequence, Bose-Einstein rather than BCS-like condensation   
 takes place.\cite{BF06,ZIBF12} This regime is beyond our weak-coupling PRM approach however.   
 
\begin{figure}[t]
    \begin{center}
      \includegraphics[angle = -0, width = 0.45\textwidth]{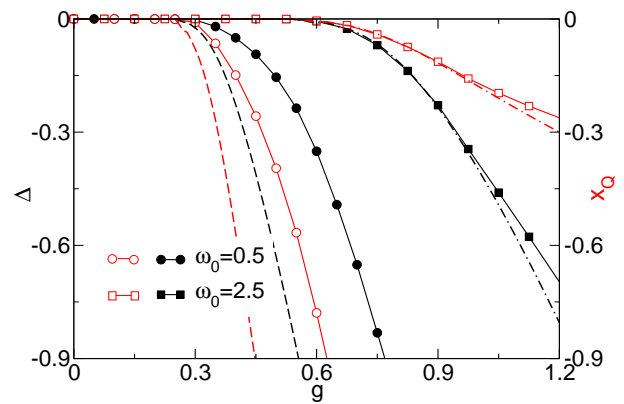}
    \end{center}
\caption{(Color online) EI order parameter $\Delta$ (black filled symbols, left axis of ordinate)  and lattice displacement, $x_Q$ 
(red open symbols, right axis of ordinate) as functions of the electron-phonon interaction $g$ in the adiabatic ($\omega_0=0.5$, circles) and 
non-adiabatic ($\omega_0=2.5$, squares) cases. Black (red) dashed [dot-dashed] lines without symbols give the corresponding mean-field results
for $\Delta$ ($x_Q$) at $\omega_0=0.5$ [$\omega_0=2.5$].}
\label{fig:D-T}
\end{figure}

 Figure~\ref{fig:D-T} now gives $\Delta$ and $x_Q$ as a function of $g$ for $\omega_0=0.5$, 2.5  and shows 
 the precision with which the critical coupling can be determined. We see that $\Delta$ and $x_Q$ are intimately related; while the semimetal 
 corresponds to an undistorted ground state, the EI/CDW state exhibits a finite lattice distortion (dimerization).  
 Actually we expect that the semimetal-EI transition is of Kosterlitz-Thouless type,\cite{KT73}  at least in the anti-adiabatic ($\omega_0\gg 1$) regime. 
If so the charge gap will open exponentially on entering the insulating phase, in line with 
 what is observed for the (repulsive Tomonaga-Luttinger liquid) metal-CDW transition in the one-dimensional spinless fermion Holstein\cite{BMH99,EF09a} and Edwards\cite{EHF09} models.   Also in conformity with the Holstein model we find that quantum phonon
 fluctuations protract the metal-insulator transition, which takes place at infinitesimal  small coupling only if $\omega_0\to 0$.  
 The effect of fluctuations/correlations beyond mean-field is obvious: within the PRM scheme a larger electron-phonon coupling $g$ is required
 to   achieve the same magnitude of the EI order or lattice displacement. The difference between PRM and mean-field results
 is insignificant, of course, for very large phonon frequencies. 

\begin{figure}[t]
    \begin{center}
      \includegraphics[angle = -0, width = 0.42\textwidth]{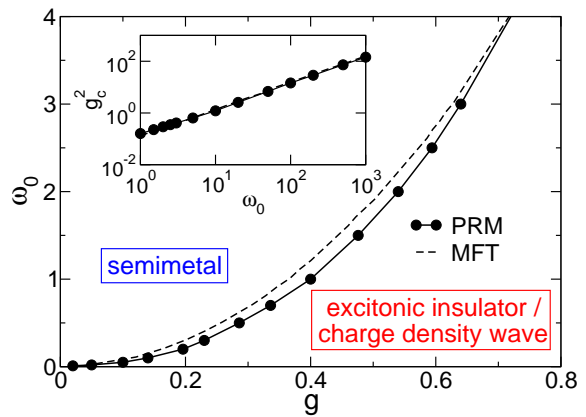}
    \end{center}
\caption{(Color online) PRM ground-state phase diagram of the two-band $f$-$c$ electron-phonon model in the $g$-$\omega_0$ plane for the half-filled band case. The inset shows the asymptotic behavior  of the (squared) critical coupling strength $g_c^2$ at very large phonon frequencies ($\omega_0\to\infty$), c.f. Appendix~B. Dashed lines give the corresponding mean-field results.}
\label{fig:pd}
\end{figure}

The derived zero-temperature quantum-phase-transition lines, separating the semimetallic and EI phases in the $g$-$\omega_0$ plane, 
are shown in Fig.~\ref{fig:pd}. In the intermediate coupling and frequency regime, we observe distinct  PRM corrections to the mean-field transition points.
In the anti-adiabatic limit ($\omega_0\to \infty$), where the phononic degrees of freedom can be integrated out, the squared critical 
coupling $g^2_c$ scales  linearly with $\omega_0$, and we find $(g^2_c/\omega_0)_{\omega_0\to\infty} \simeq 0.14$ (see inset). 
Here the phase boundary basically  agrees with that of mean-field theory (compare Fig.~\ref{fig:pd}).
 An analytical proof of this finding is given in Appendix~\ref{B}.   

\subsection{Spectral properties} 
\label{S:V.B}

\begin{figure}[t]
    \begin{center}
      \includegraphics[angle = -90, width = 0.218\textwidth]{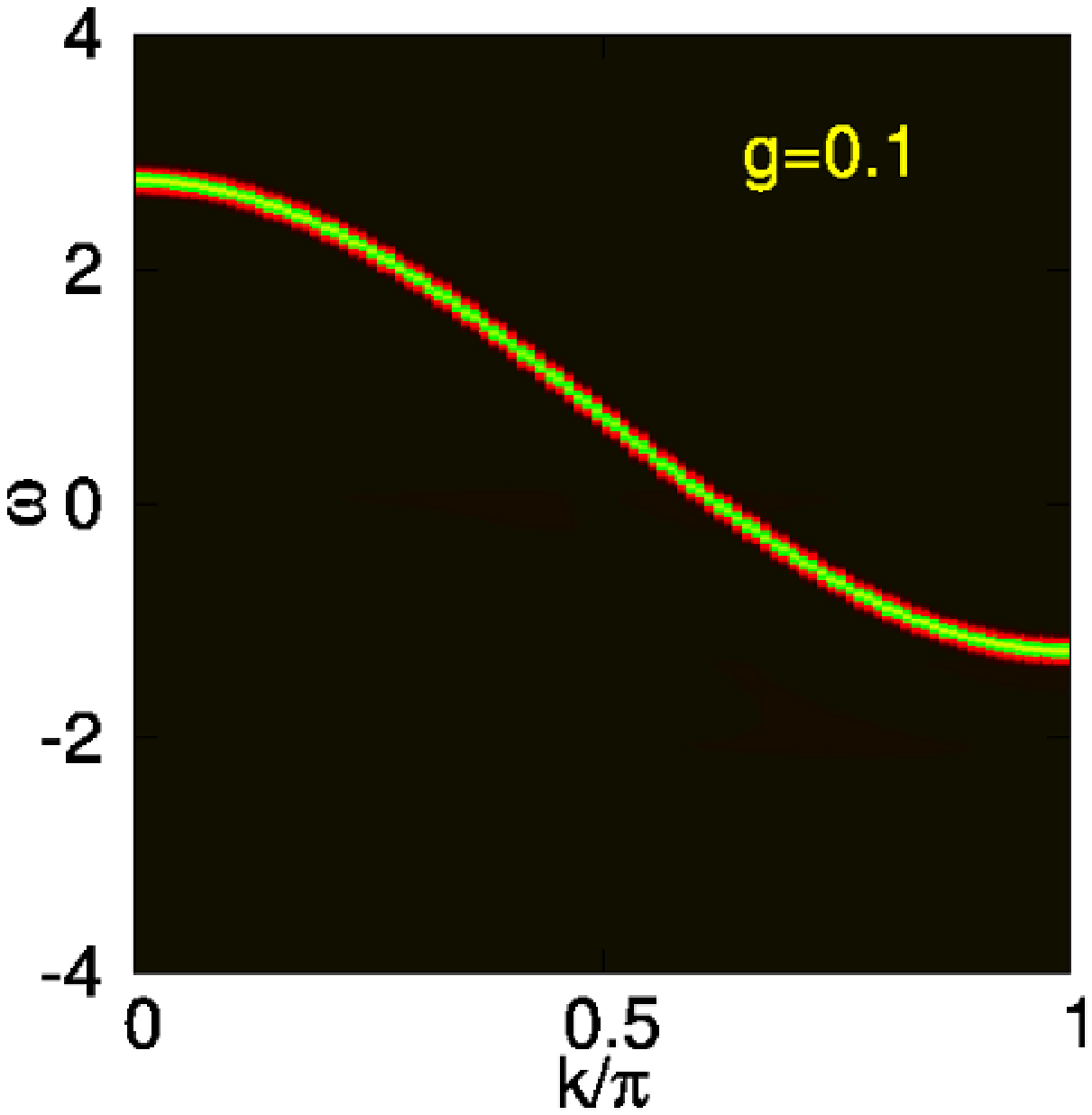}
      \includegraphics[angle = -90, width = 0.249\textwidth]{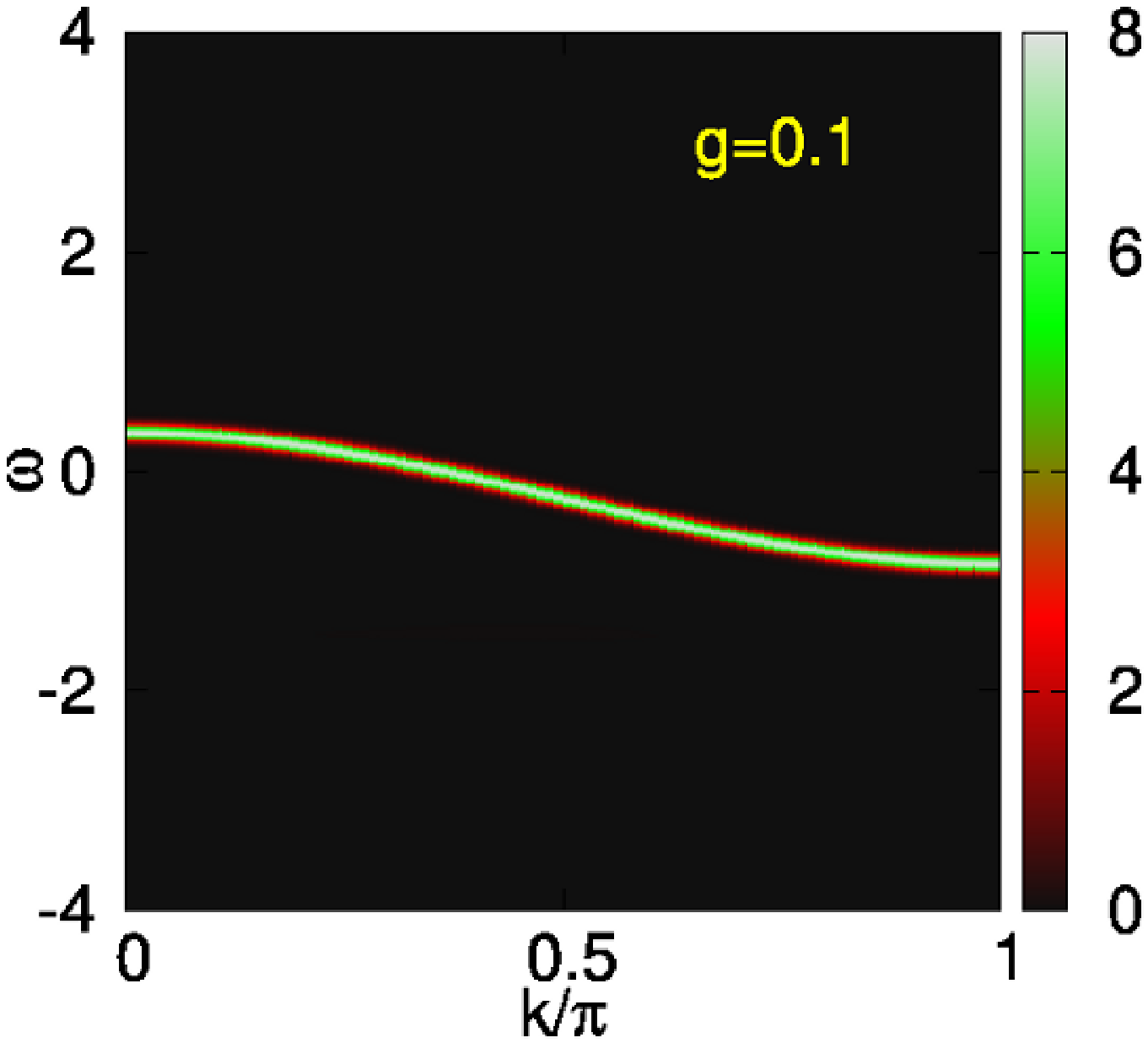}
      \includegraphics[angle = -90, width = 0.218\textwidth]{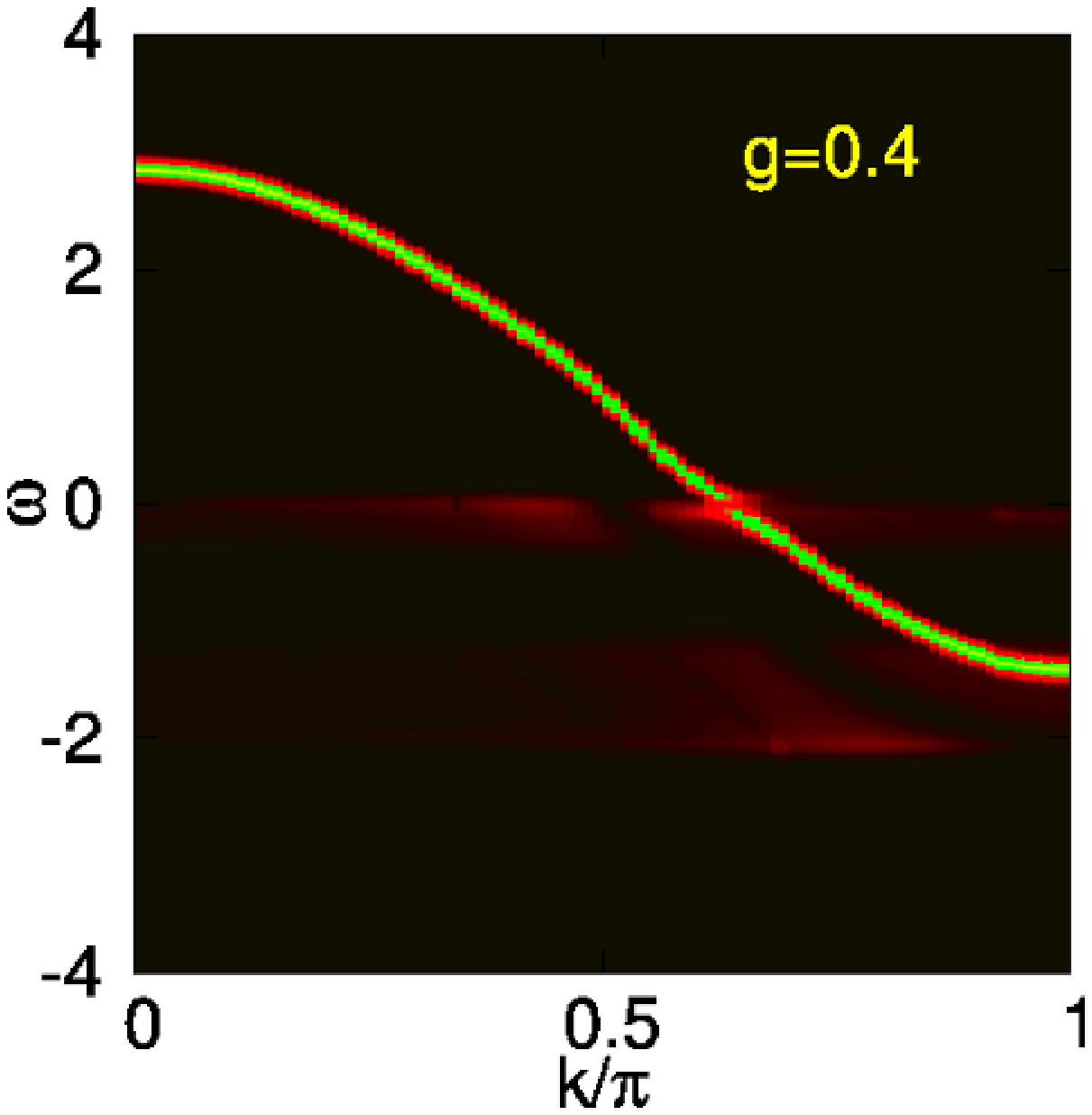}
      \includegraphics[angle = -90, width = 0.249\textwidth]{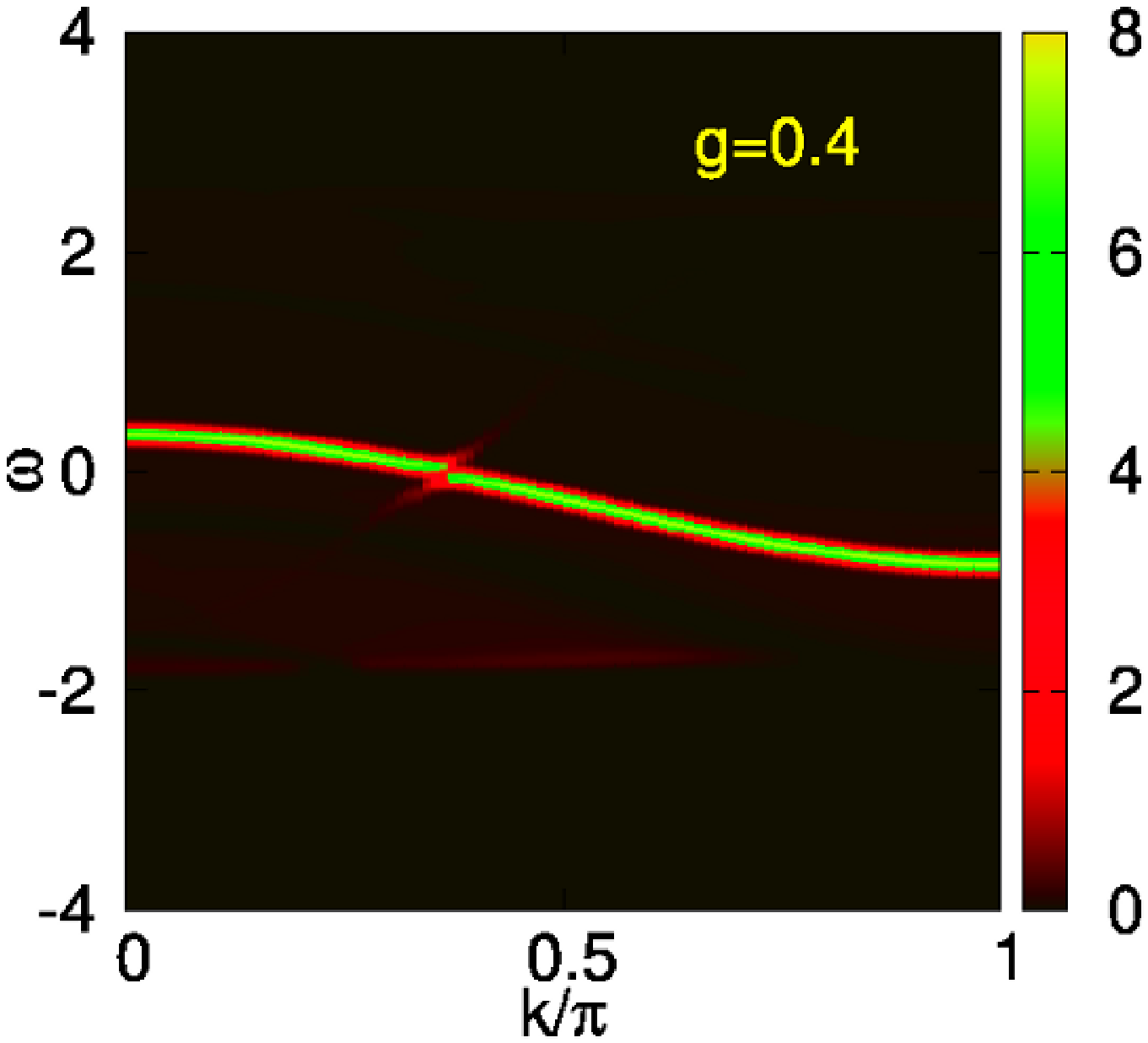}
      \includegraphics[angle = -90, width = 0.218\textwidth]{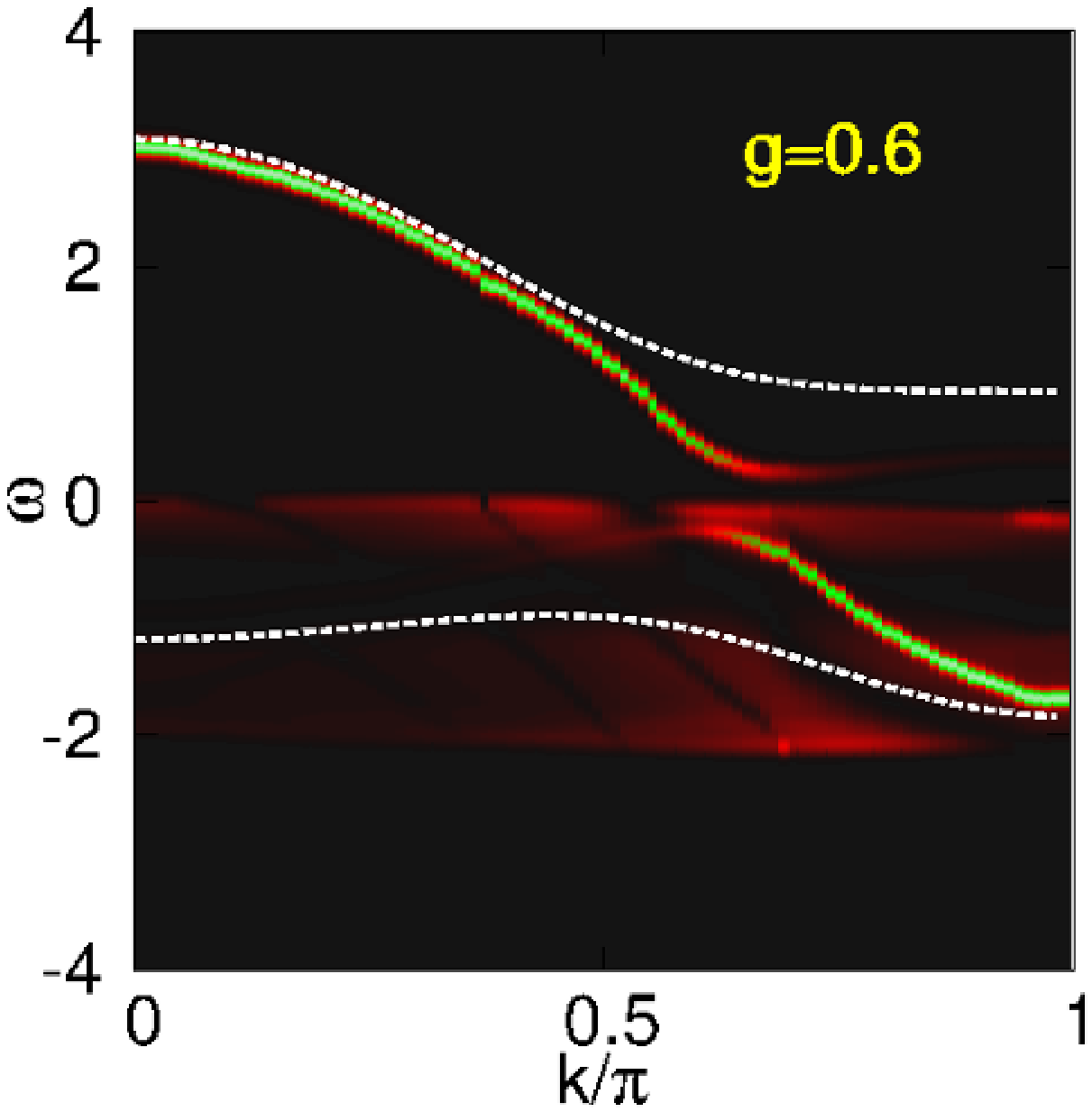}
      \includegraphics[angle = -90, width = 0.249\textwidth]{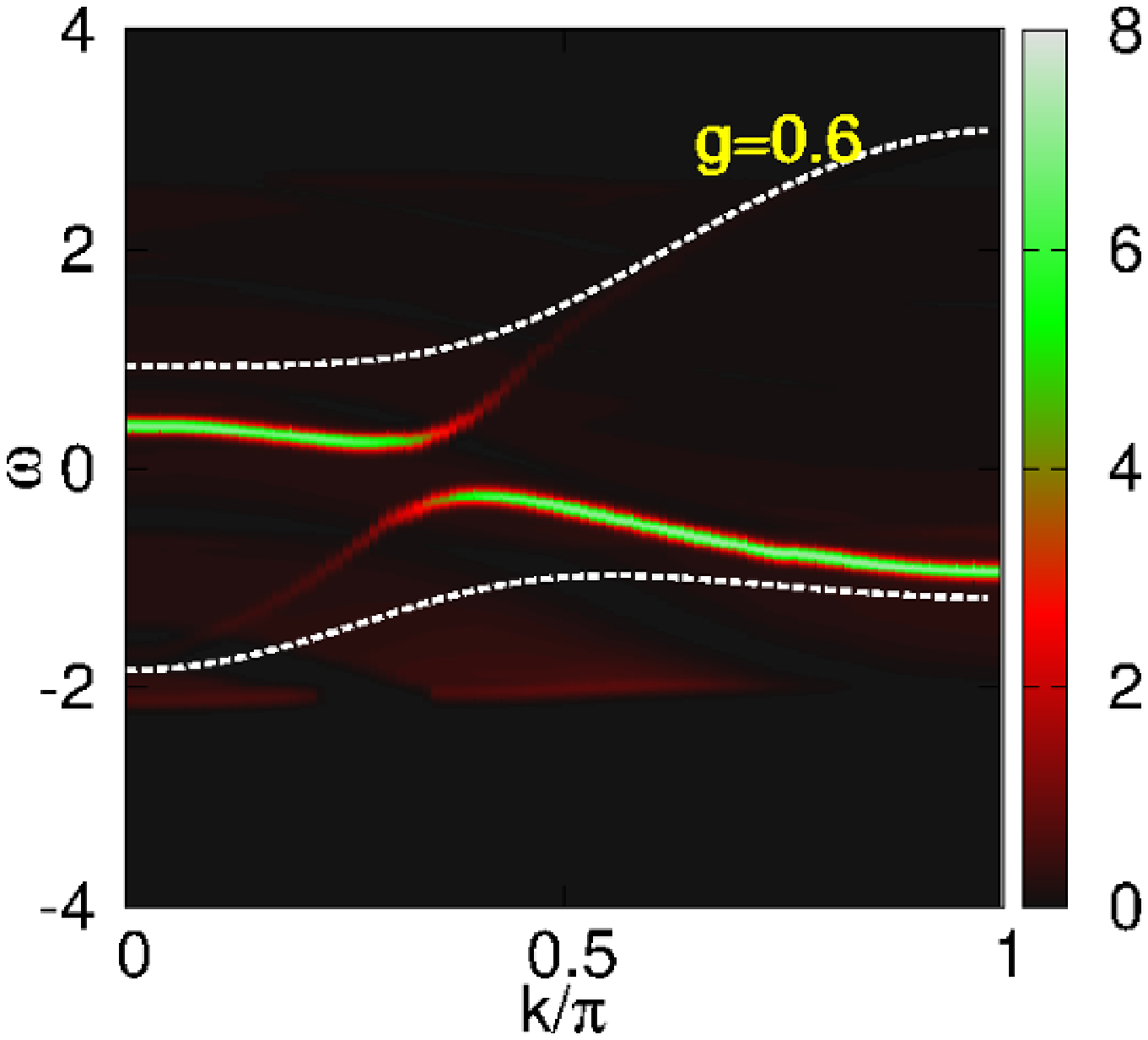}
    \end{center}
\caption{(Color online) Intensity plots of the $c$-electron (left-hand panels) and $f$-electron (right-hand panels) 
single-particle spectral functions $A^{c,f}_{k}(\omega)$ in the adiabatic regime with $\omega_0=0.5$. The electron-phonon 
coupling $g$ increases as indicated from top to bottom panels. In the lowermost two panels the corresponding mean-field results are included, 
without dissolving the spectral intensity however (see white dashed lines).} 
\label{fig:Acf1}
\end{figure}

\begin{figure}[t]
    \begin{center}
      \includegraphics[angle = -90, width = 0.218\textwidth]{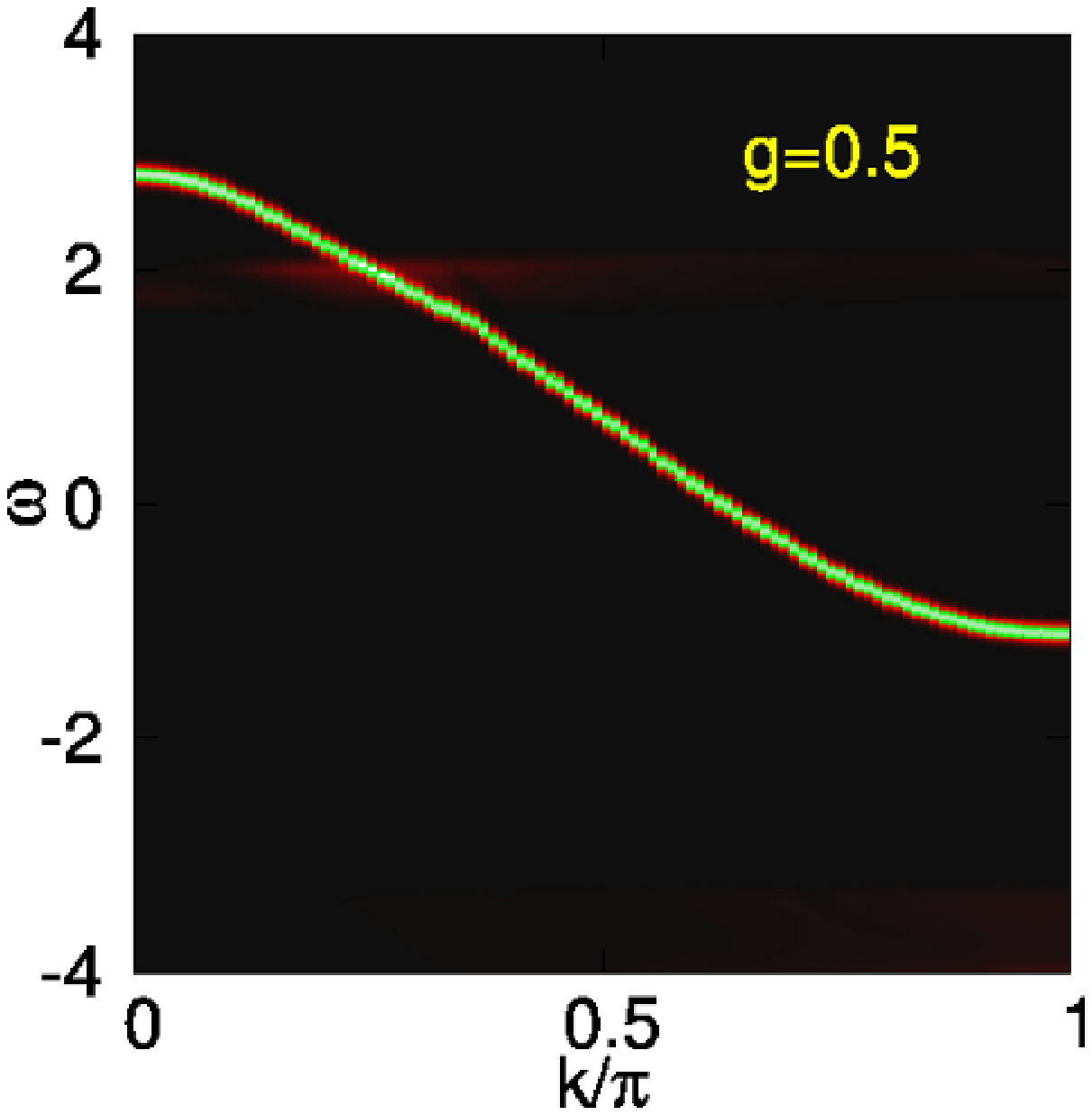}
     \includegraphics[angle = -90, width = 0.249\textwidth]{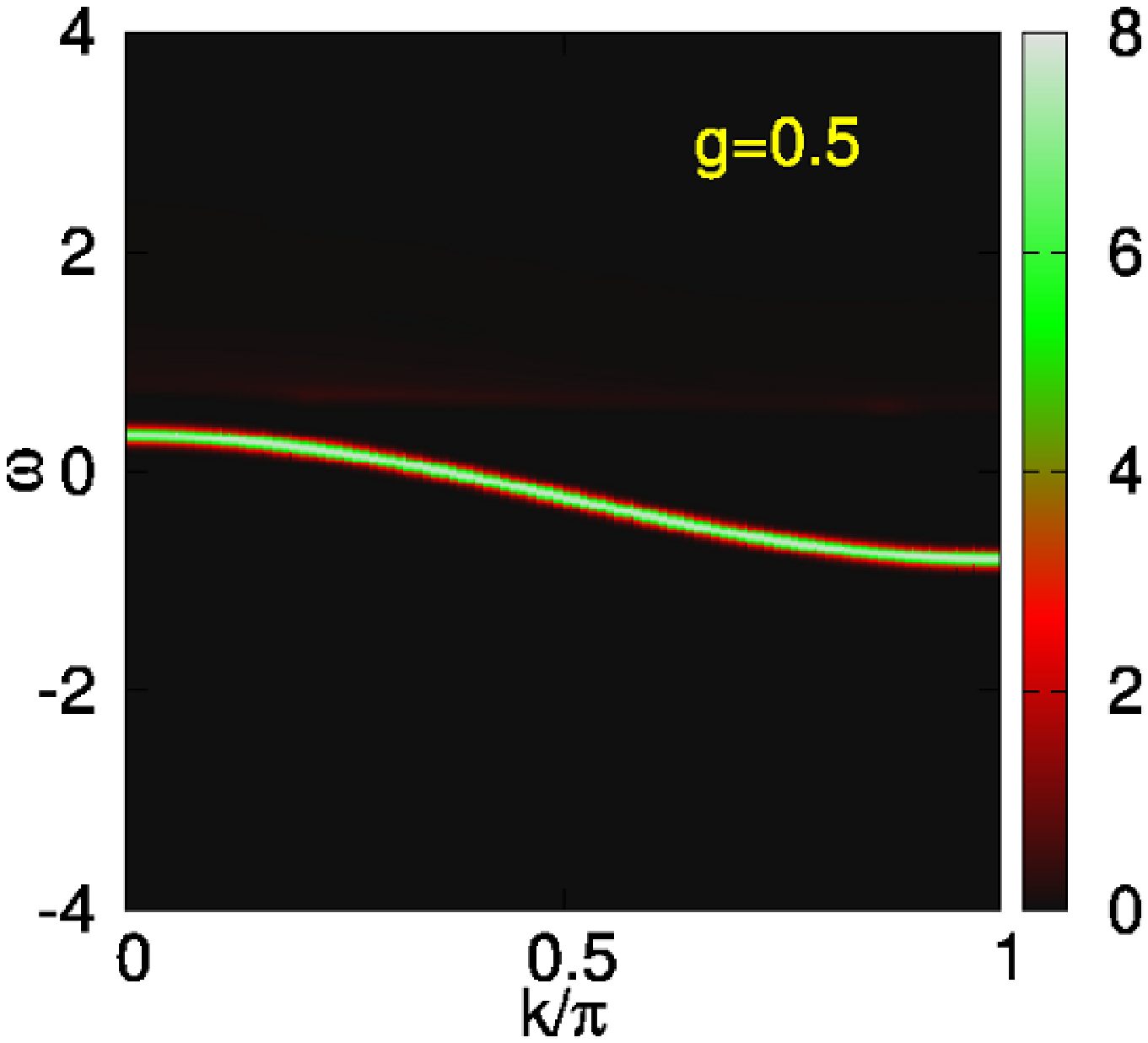}
      \includegraphics[angle = -90, width = 0.218\textwidth]{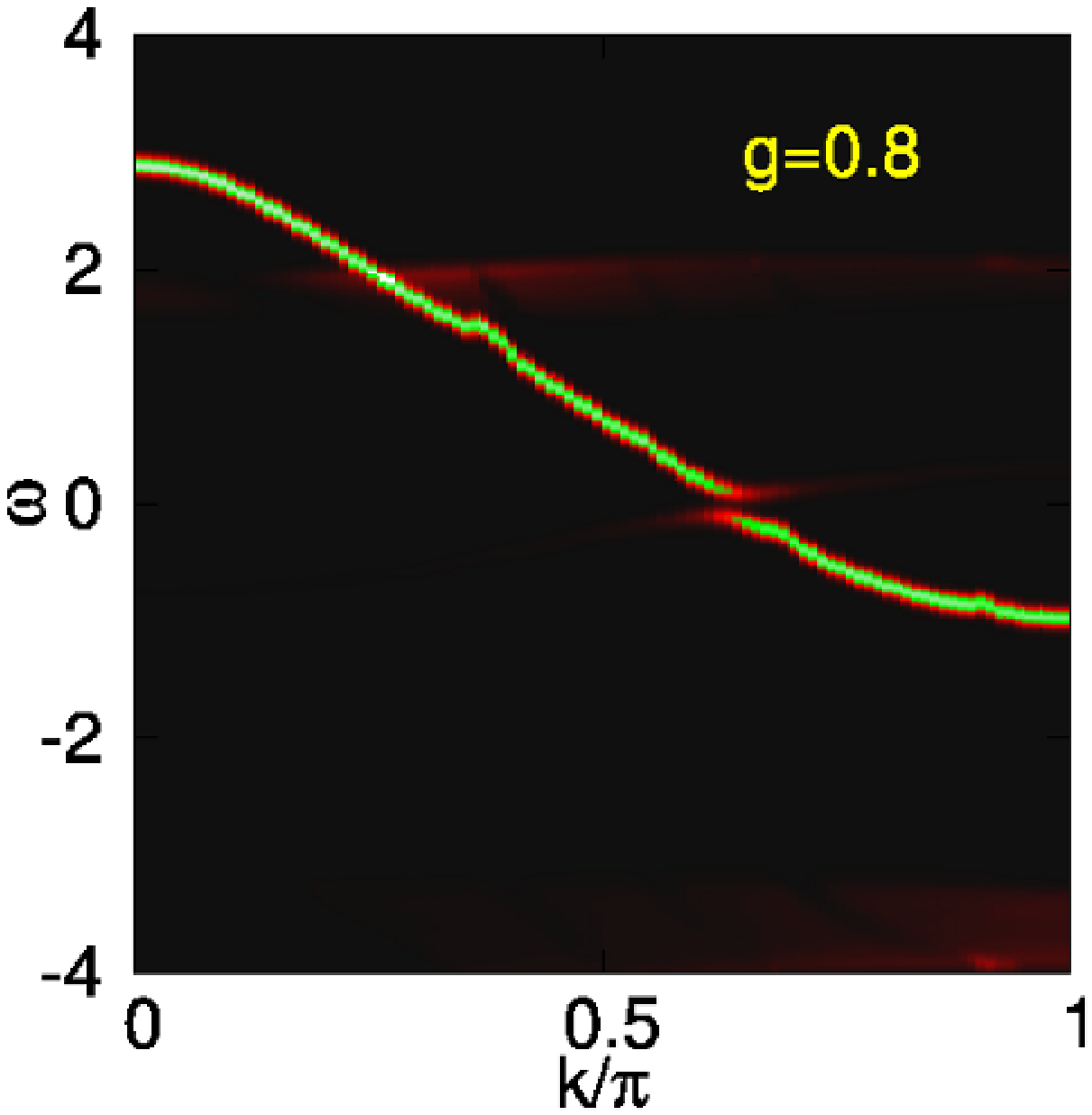}
       \includegraphics[angle = -90, width = 0.249\textwidth]{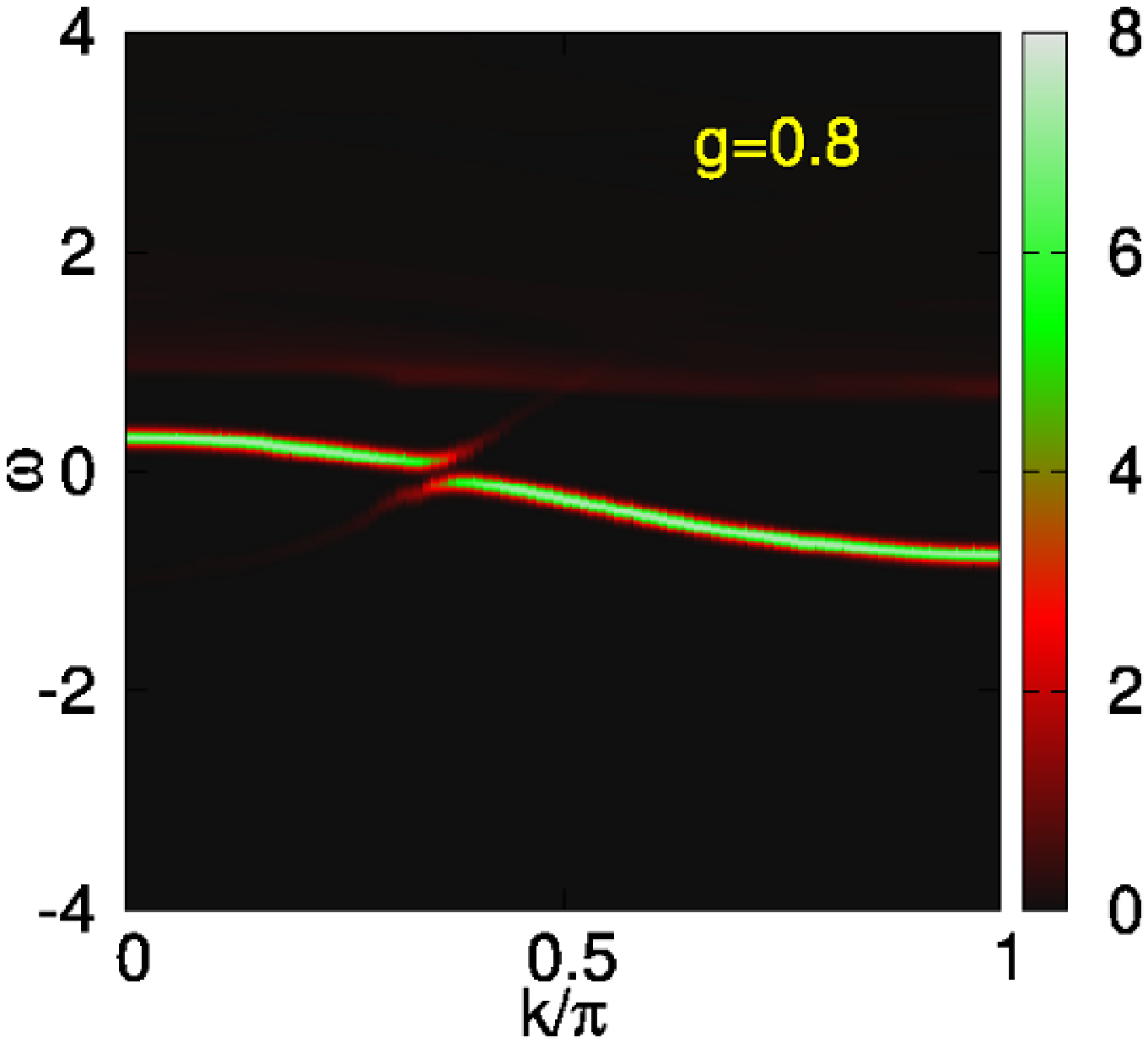}
    \end{center}
\caption{(Color online) Intensity plots of the $c$-electron (left-hand panels) and $f$-electron (right-hand panels) 
single-particle spectral functions $A^{c,f}_{k}(\omega)$ for $g=0.5$ (top panels) and  $g=0.8$ (bottom  panels) in the non-adiabatic regime with $\omega_0=2.5$.}
\label{fig:Acf2}
\end{figure}

We now present the PRM results for the single-particle spectral functions  associated with the photoemission or inverse photoemission
(injection) of a $c$ or $f$ electron with wave number  $k$ and energy $\omega$, which serve as a direct measures of the occupied and unoccupied states. Figure~\ref{fig:Acf1} shows the variation of $A^{c,f}_{k}(\omega)$ as the electron-phonon coupling increases in the adiabatic regime. 
For weak couplings (see upper panels), we are in the semimetallic phase, and $A^{c,f}_{k}(\omega)$ reflect the weakly renormalized $c$- and $f$-band dispersions (note that the energy $\omega$ is measured with respect to the  Fermi energy). 
In the EI phase, a gap opens at the Fermi energy and we observe a pronounced back-folding of the spectral signature at larger coupling. 
Here $c$- and $f$-electron states strongly hybridize close to the Fermi energy. The same, in principle, holds in the non-adiabatic regime.
However,  for the parameters used in Fig.~\ref{fig:Acf2}, the ratio $g/\omega_0=0.32$ is much smaller than for the EI phase depicted in
the two lowermost panels of Fig.~\ref{fig:Acf1} where $g/\omega_0=1.2$. Hence multi-phonon processes are less important in the former case and---prescinding from the gap feature---the photoemission spectrum is less affected by the lattice degrees of freedom. For comparison, we show also the outcome of mean-field theory for  
$A^{c}_{k}(\omega)$ and $A^{f}_{k}(\omega)$ in  lower panels where $g=0.6$. We see that the band gap is considerably overestimated, and there is, of course, no 
incoherent contribution at all. 
\begin{figure}[h]
    \begin{center}
      \includegraphics[angle = -90, width = 0.218\textwidth]{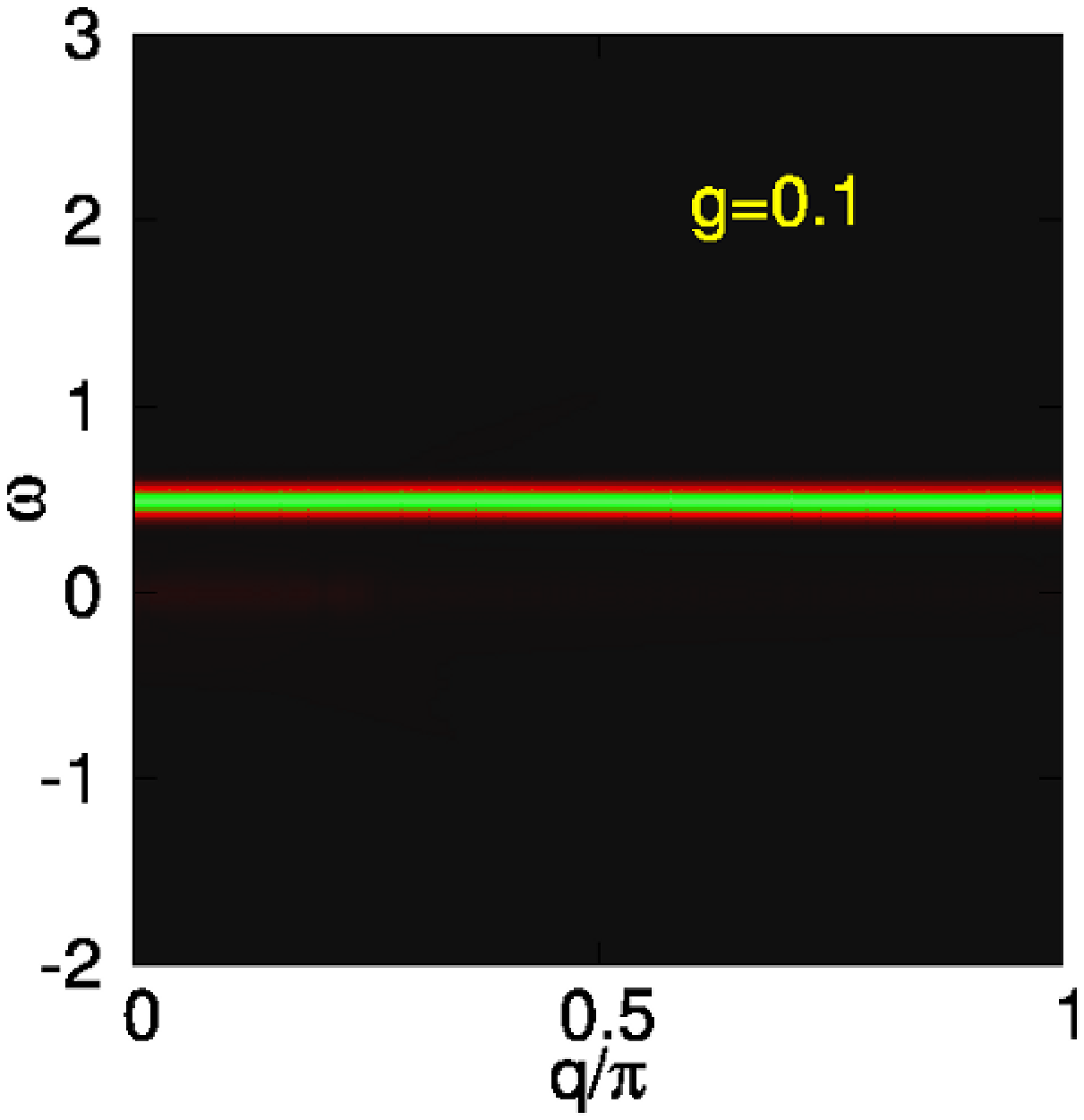}
      \includegraphics[angle = -90, width = 0.255\textwidth]{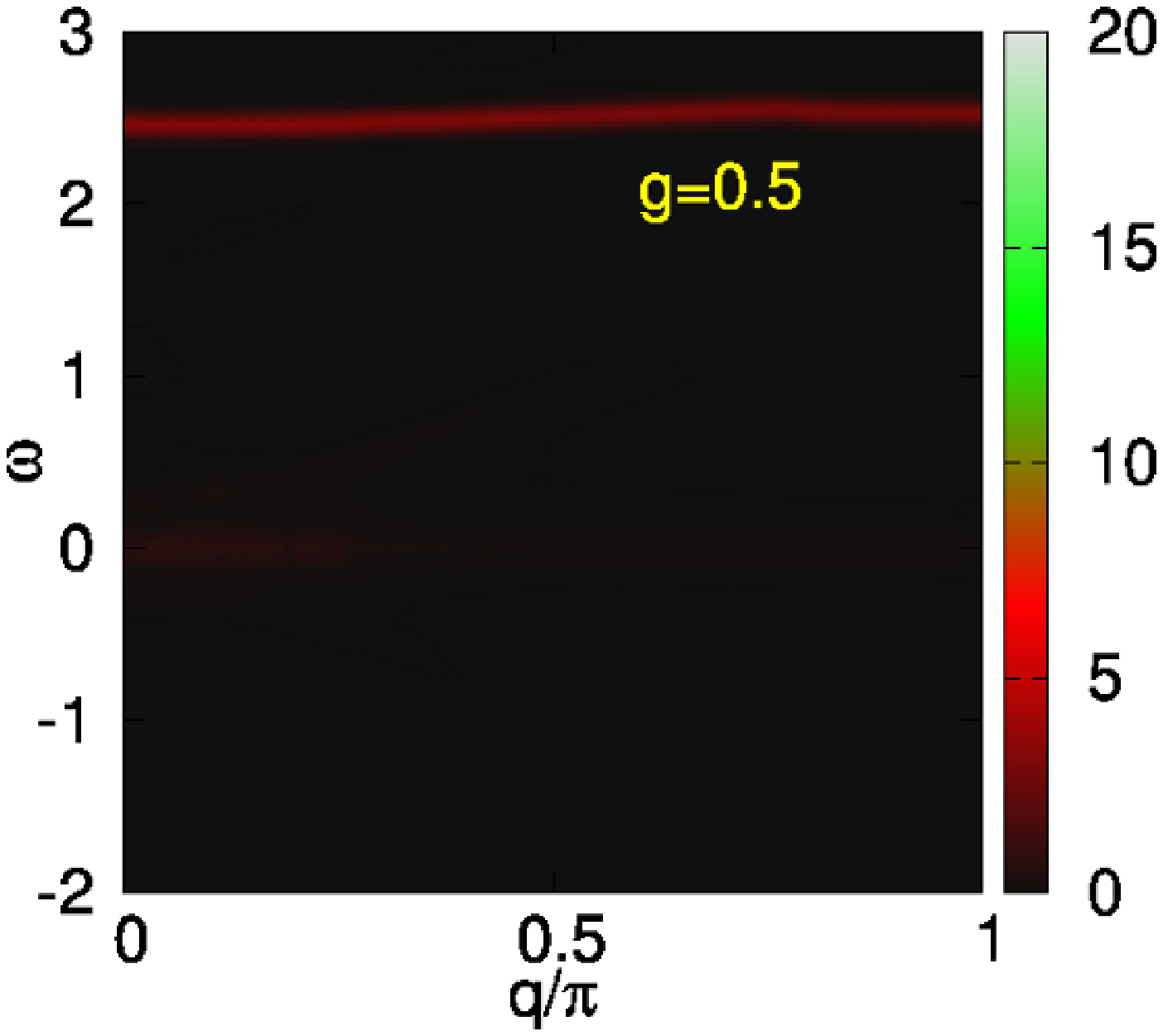}
      \includegraphics[angle = -90, width = 0.218\textwidth]{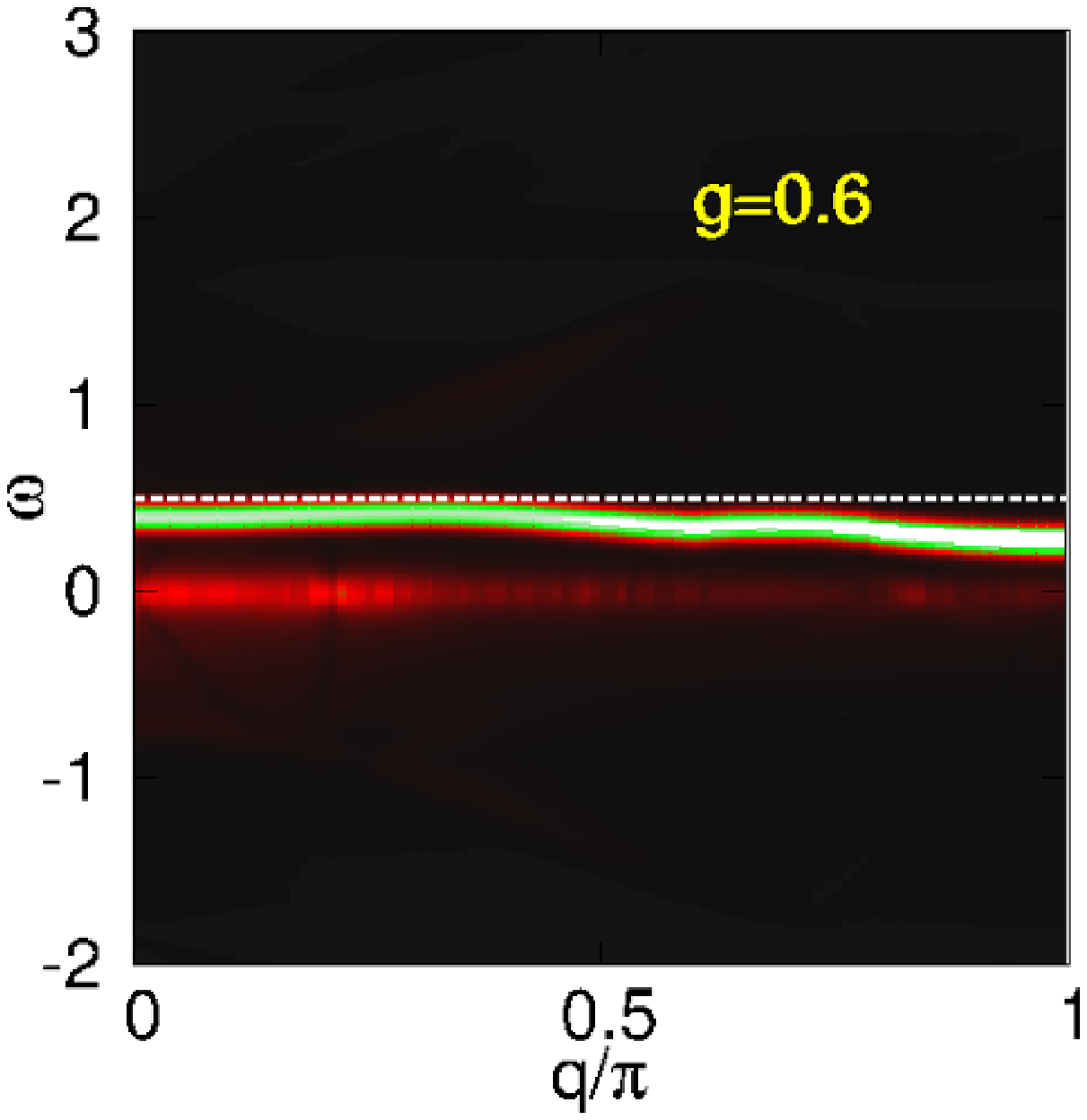}
      \includegraphics[angle = -90, width = 0.25\textwidth]{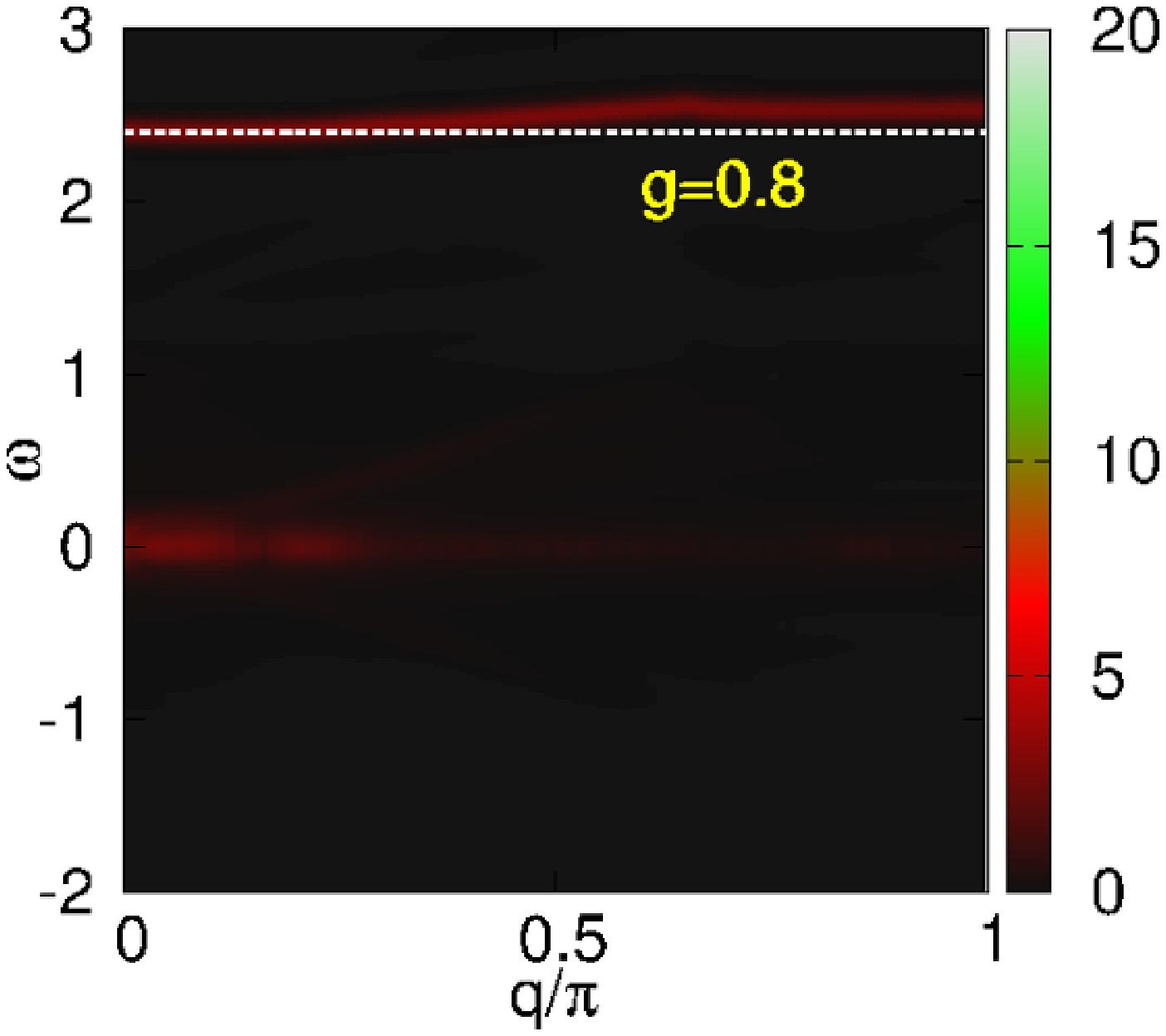}
    \end{center}
\caption{(Color online) Intensity plots of the phonon spectral function $C_{q}(\omega)$ for different $g$ at $\omega_0=0.5$
(left-hand panels) and $\omega_0=2.5$, right-hand panels). The straight white dashed line in the lower panels marks the dispersionless mean-field result.}  \label{fig:Ap}
\end{figure}

More information in this respect comes from the phonon spectral function $C_{q}(\omega)$, represented in Fig.~\ref{fig:Ap}, below (upper panels) and above (lower panels) the semimetal-EI transition point.  At weak coupling, the absorption signal is dominated by 
the coherent part of $C_{q}(\omega)$, which is almost dispersionsless and located near the bare phonon frequency,
i.e., $\tilde{\omega}_{q}\simeq \omega_0$. This particularly holds for the case $g=0.1$, $\omega_0=0.5$ 
shown in the  top left panel.
For $g=0.5$ and a higher phonon frequency $\omega_0=2.5$, the overall intensity of the signal goes down, of course. 
Note that the phonon mode acquires a slight dispersion: It becomes larger near the Brillouin zone boundary ($\tilde{\omega}_{\pi}\gtrsim \omega_0$). Above the transition [$g>g_c(\omega_0$)],  we observe two distinct features [see lower panels of Fig.~\ref{fig:Ap}].  Firstly, the phonon mode softens for $\omega_0=0.5$ while it hardens for $\omega_0=2.5$. That is, we find an opposite tendency for small and large phonon frequencies.  
This results can already be understood from perturbation theory for the phonon energy
as shown in Appendix \ref{A:4}.
Secondly a new signal at $\omega=0$ appears which indicates the strong coupling between electronic and phononic degrees of freedom. 
Note that the phonon spectral function calculated within mean-field approximation shows only a single dispersionless signal at $\omega=\omega_0$.

\begin{figure}[t]
    \begin{center}
      \includegraphics[angle = -90, width = 0.23\textwidth]{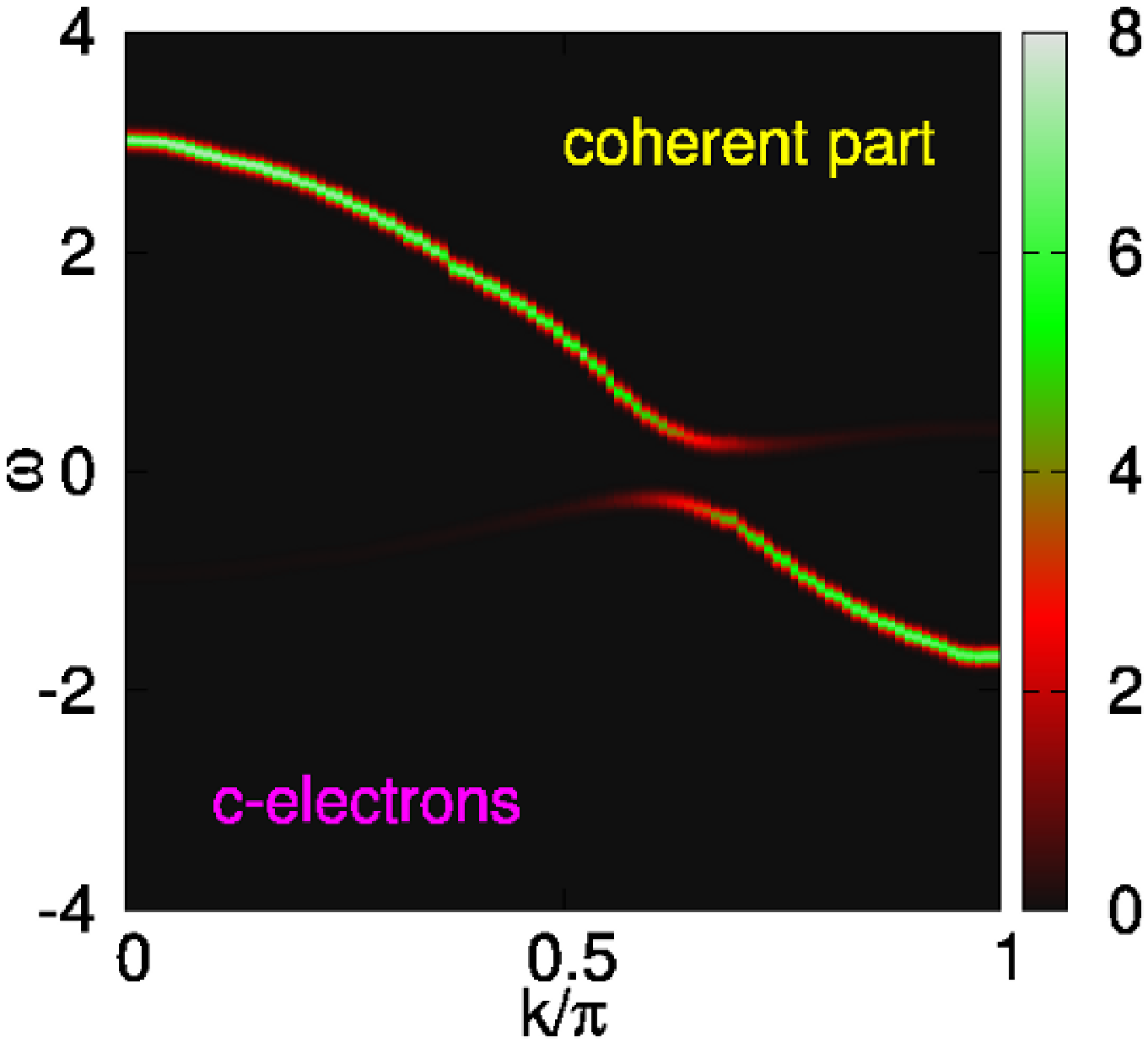}
      \includegraphics[angle = -90, width = 0.24\textwidth]{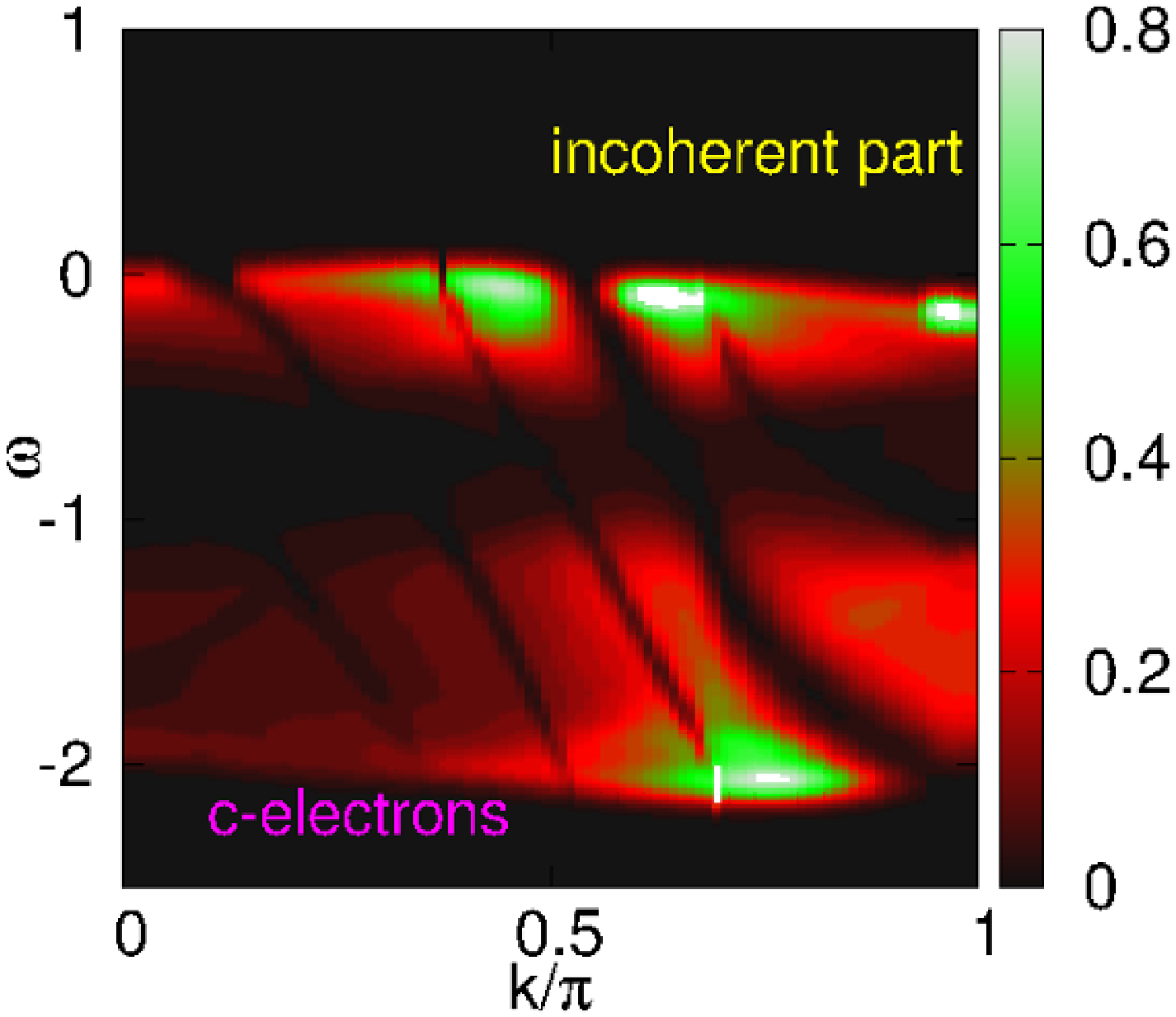}
      \includegraphics[angle = -90, width = 0.23\textwidth]{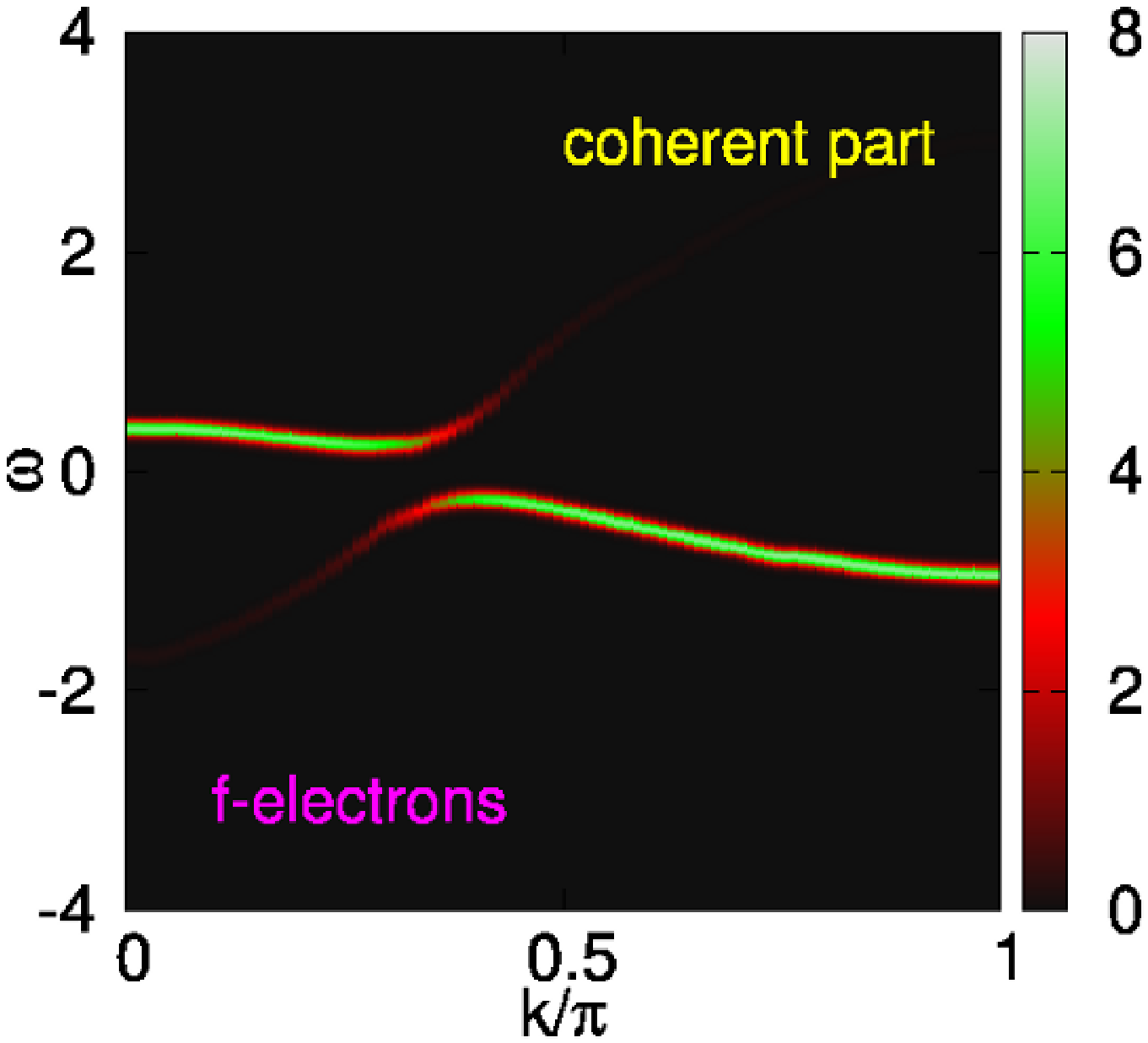}
      \includegraphics[angle = -90, width = 0.24\textwidth]{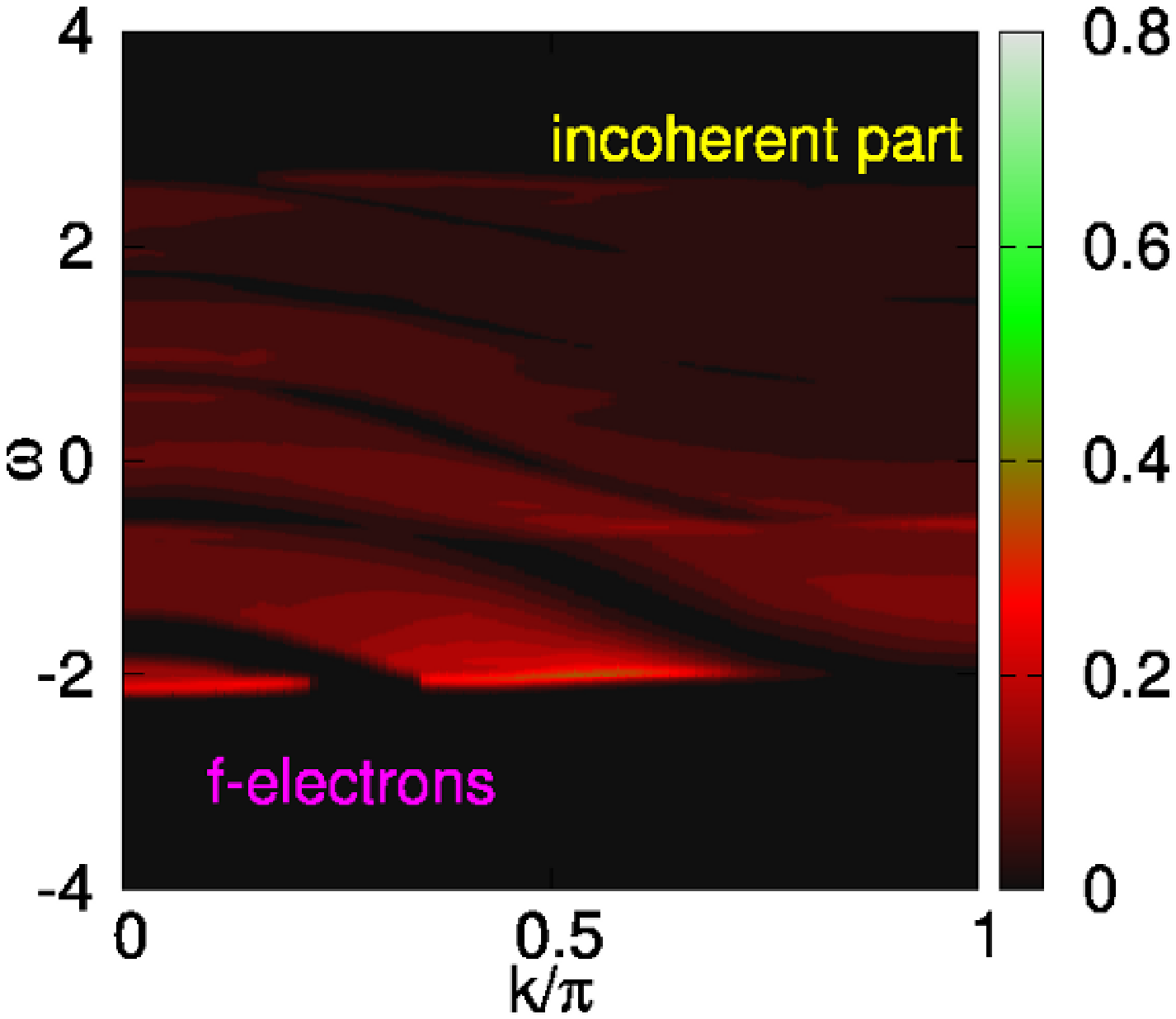}
      \includegraphics[angle = -90, width = 0.23\textwidth]{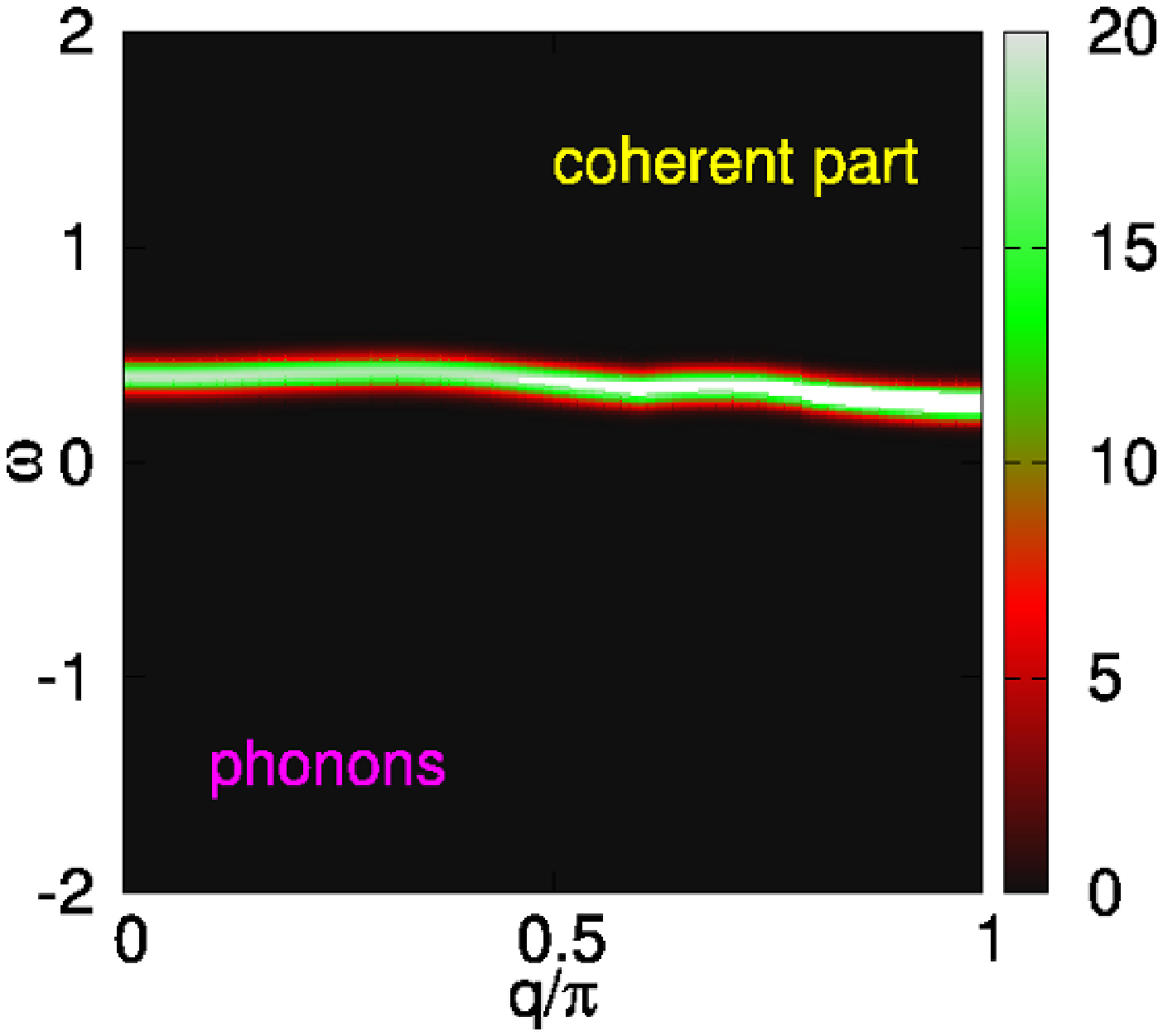}
      \includegraphics[angle = -90, width = 0.235\textwidth]{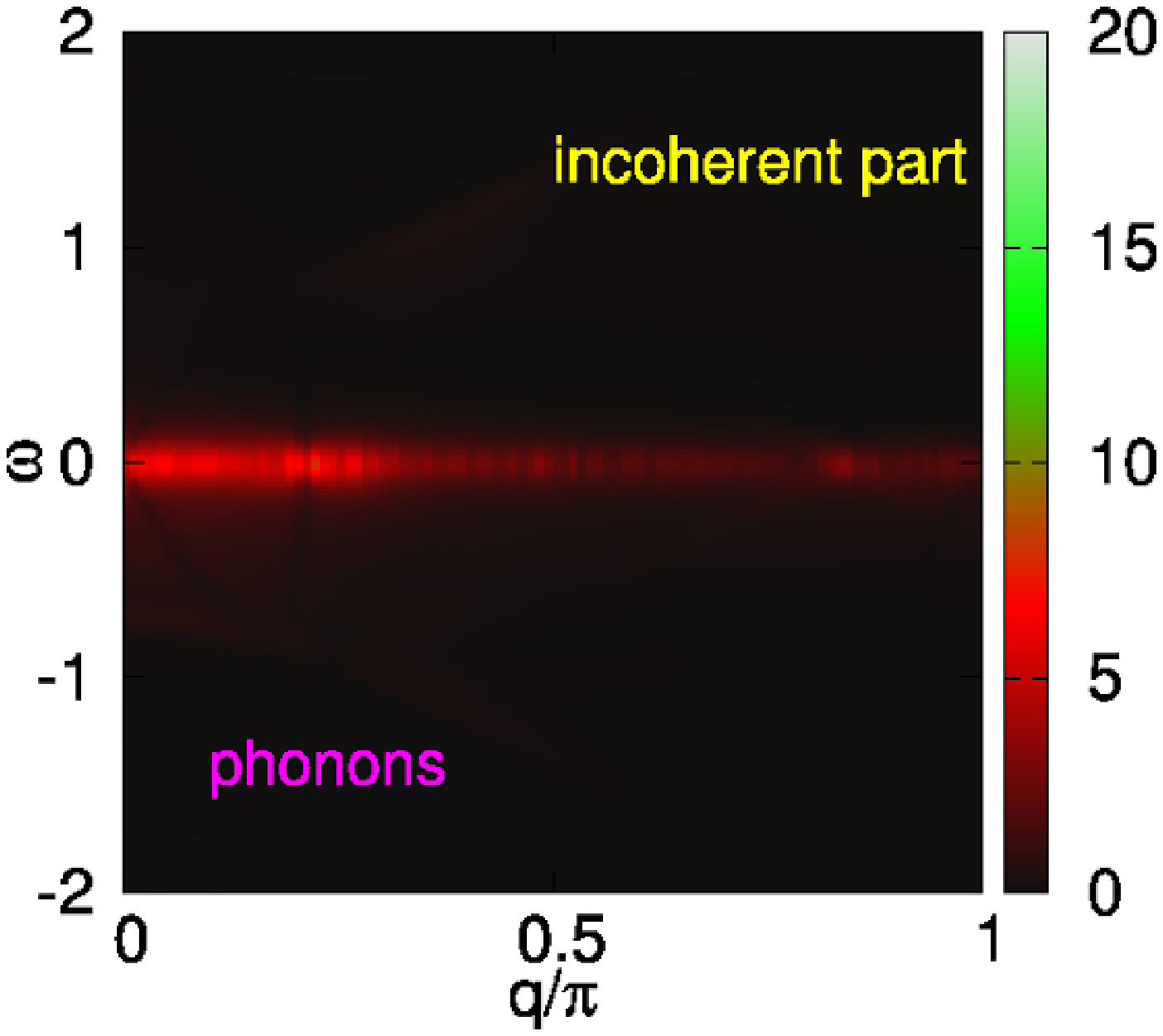}
    \end{center}
\caption{(Color online) Intensity plots of the coherent (left-hand panels) and the incoherent (right-hand panels) parts of the 
$c$- and $f$-electron single particle spectral functions and of the phonon spectral function. Note the different color coding
of the coherent and incoherent parts of $A^{c,f}_{k}(\omega)$. We stress that also $A^{c}_{k}(\omega)$
has a finite incoherent part for $\omega >0$ (as magnification would show), which only is noticable  in a small range above $\omega =0$ however, because--amongst others--the renormalized 
$f$ bandwidth is small. Results are  given for 
$\omega_0=0.5$ and $g=0.6$.}
\label{fig:sw}
\end{figure}

Besides coherent excitations  all spectral functions in Figs.~\ref{fig:Acf1}-\ref{fig:Ap} 
also show incoherent excitations. They can be detected as (red-colored) much 
weaker developed contributions which deviate from the coherent ones. They possess
two general features: (i) Their weight increases with increasing electron-phonon 
coupling $g$ since they are induced by $\mathcal H_1$, 
and (ii) their weight is strongly suppressed in the anti-adiabatic limit.  This is explained in Appendix \ref{B}. 
To elucidate the distribution of the spectral weight in more detail, we show in Fig.~\ref{fig:sw} the coherent (left) and incoherent (right)
part of the  $A^{c}_{k}(\omega)$, $A^{f}_{k}(\omega)$ and  $C_{q}(\omega)$  spectral functions separately. We choose as  an example $g=0.6$ and $\omega_0=0.5$, i.e., consider the system to be in the (adiabatic) EI/CDW regime (cf. Fig.~\ref{fig:pd}).   For these parameters the coherent signatures, given by the first terms in Eqs.~\eqref{A49}, \eqref{A50} and~\eqref{A51}  clearly dominate the spectra in each case (note that the intensity of the incoherent contributions is magnified by a factor of ten). 
They follow the renormalized dispersions $E_k^c$ and $E_k^f$, possessing an excitation gap  and a pronounced $c$-$f$ electron 
hybridization. Obviously the incoherent contribution of the $c$-electron spectrum is noticeable in the range of 
the $f$-electron band  only (and vice versa) and will be enhanced if  electron-phonon coupling increases. 
The phonon spectral function reveals that the signal at $\omega\simeq 0$ originates from the incoherent part of the spectrum. It acquires substantial 
spectral weight only if the renormalized quasiparticle bands will be `connected' by phonon absorption/emission processes which are significant at large $g$.   

\section{Summary}
\label{S:VI}
Applying  a discrete version of the projective renormalization method to a two-band $f$-$c$ electron model 
with coupling to the lattice degrees of freedom we show that the exciton-phonon interaction can drive a 
semimetal--to--excitonic insulator transition at zero temperature in one dimension. The ground-state phase diagram 
containing semimetallic and excitonic insulator phases is derived. The excitonic condensate 
is accompanied by a charge density wave and a finite lattice dimerization, and is intimately connected with
a developing $f$-$c$ electron hybridization/coherence. At finite phonon frequency, this spontaneously symmetry-broken  state does not appear until the interaction exceeds a finite critical coupling strength. The phase boundary determined by the projective renormalization method  significantly deviates from the mean-field result in the intermediate exciton-phonon coupling and phonon frequency regime. The quantum phase transition shows up in the spectral quantities  too: We notice the opening of a single-particle excitation gap in the photoemission spectrum, a substantial spectral weight transfer from the coherent to the incoherent part of the spectra,  and a renormalisation of the phonon mode, which becomes softened (hardened) as the transition point is reached in the adiabatic (non-adiabatic to anti-adiabatic) regime. In this way our work points out the prominent role played by the lattice degrees of freedom establishing 
a charge density wave in semimetallic systems with weak (indirect) band overlap and in mixed-valent semiconductors with band gaps 
comparable to the exciton binding energy, such as quasi two-dimensional 1$T$-$\rm TiSe_2$ and three-dimensional $\rm TmSe_{0.45}Te_{0.55}$, respectively. 
For 1$T$-$\rm TiSe_2$ it has been shown quite recently in the framework of a multiband extended Falicov-Kimball model that a purely electronic, exciton pairing and condensation mechanism is insufficient to describe the observed (long-ranged) chiral charge order.\cite{ZFBMB13} Hence the coupling to the phonons seems to be essential, and first mean-field results 
indicate that electron-hole Coulomb attraction and exciton-phonon coupling indeed support each other in establishing a charge-density-wave state with small but finite lattice distortion. Thereby essential electron 
correlation and all phonon fluctuation effects were neglected however.  Here we consider a one-dimensional  model, where quantum phonon fluctuations are exceedingly 
important and in general tend to suppress any long-range charge order or lattice dimerization, and show that an electron/hole-lattice coupling of reasonable strength can 
nonetheless  cause an excitonic instability. A future more complete theoretical discussion of the low-temperature properties of these material classes 
should definitely comprise the complex interplay of electron-phonon and electron-hole interactions beyond mean field, 
particularly in the vicinity of the semiconductor-semimetal transition.

\acknowledgements
The authors would like to thank H. Beck, F. X. Bronold,  S. Ejima, D. Ihle, S. Sykora, and B. Zenker 
for valuable discussions. This research is funded by Vietnam National Foundation for Science and Technology
Development (NAFOSTED) under grant number 103.02-2012.52 and by Deutsche Forschungsgemeinschaft through SFB 652, B5.


\begin{appendix} 

\section{Projector-based Renormalization Method}
\label{A}

In this Appendix we demonstrate in detail how  to solve Hamiltonian $\mathcal H$ 
by means of the PRM. So far he PRM was successfully applied to the one-dimensional Holstein\cite{SHBWF05} and 
extended Falicov-Kimball\cite{PBF10}  models and a number of other models. Its starting point is the decomposition 
of a many-particle Hamiltonian $\mathcal H$ into an `unperturbed' part  
$\mathcal{H}_0$ and into a `perturbation' $\mathcal{H}_1$, 
where the unperturbed part $\mathcal{H}_0$ is  solvable.  
The perturbation is responsible for transitions between the eigenstates of 
$\mathcal H_0$ with non-vanishing transition energies $|E^n_0 -E^m_0|$.  Here $E^n_{0}$ and $E^m_{0}$ 
denote the energies of $\mathcal H_{0}$ between which the transitions take place.
The basic idea of the PRM method is to integrate out the interaction $\mathcal{H}_1$ 
by a sequence of discrete unitary transformations\cite{BHS02}. 
Thereby, the PRM renormalization starts from the largest transition energy of the original Hamiltonian $\mathcal H_0$, 
which will be called  $\Lambda$, and proceeds in steps $\Delta \lambda$ to lower values of 
transition energies $\lambda$.  
For practical applications the unitary transformations are best done 
in small steps $\Delta \lambda$. Thereby, the evaluation in each step can be restricted to low orders in $\mathcal H_1$. 
This procedure usually limits the validity of the approach to parameter values of $\mathcal H_1$ 
which are of the same magnitude as those of $\mathcal H_0$. 
 Every renormalization 
step is performed by means of a small unitary transformation, where all excitations between 
$\lambda$ and $\lambda - \Delta \lambda$ are eliminated:
\begin{eqnarray}
 \label{A1} 
 \mathcal H_{\lambda -\Delta \lambda} &=& e^{X_{\lambda, \Delta \lambda}} \, \mathcal H_\lambda \,
 e^{-X_{\lambda, \Delta \lambda}} \, .
\end{eqnarray}
Here, the operator  $X_{\lambda, \Delta \lambda} =- X_{\lambda, \Delta \lambda}^\dag$ 
is the generator of the unitary transformation for the small step. 
After each step both the unperturbed part
of the Hamiltonian and the perturbation become renormalized and depend on the cutoff energy $\lambda$, i.e.~one
arrives at a renormalized Hamiltonian $\mathcal H_\lambda = \mathcal H_{0,\lambda} + \mathcal H_{1,\lambda}$.
Note that $\mathcal H_{1,\lambda}$ now only accounts for transitions with energies smaller than $\lambda$.  
Proceeding the renormalization stepwise up to zero transition energies $\lambda=0$ all transitions 
with energies different from zero have been integrated out. Thus, finally one arrives 
at a  renormalized Hamiltonian $\mathcal H_{\lambda=0}$, which is diagonal or at least quasi-diagonal, since all 
transitions from $\mathcal H_1$ have been used up.

\subsection{Hamiltonian $\mathcal H_\lambda$}
\label{A:1}

Assuming that all transitions with energies larger than $\lambda$ are already integrated out, 
an appropriate {\it ansatz} in the present case for the transformed Hamiltonian $\mathcal H_\lambda$ reads, 
$\mathcal H_\lambda= \mathcal H_{0,\lambda} + \mathcal H_{1,\lambda}$ with  
 \begin{align}
 \label{A2}
\mathcal{H}_{0, \lambda}&=\sum_{\mathbf{k}}{\varepsilon}^f_{\mathbf{k}, \lambda}f^\dagger_{\mathbf{k}}f^{}_{\mathbf{k}}
+\sum_{\mathbf{k}}{\varepsilon}^c_{\mathbf{k}, \lambda}c^\dagger_{\mathbf{k}}c^{}_{\mathbf{k}}
+ \sum_{\mathbf{q}} \omega_{\bf q, \lambda}B^\dagger_{\mathbf{q}, \lambda} B^{}_{\mathbf{q, \lambda}}\nonumber\\
&+  \Delta_\lambda \sum_{\bf k} \big( c^\dag_{\bf k +\bf Q} f_{\bf k} + f^\dag_{\bf k} c_{\bf k + \bf Q} \big)\,,
\\ \label{A3} 
 \mathcal H_{1, \lambda} &=
\frac{g}{\sqrt{N}}\sum_\mathbf{kq}\mathbf P_\lambda \Big[ \delta(c^\dagger_{\mathbf{k+q}}f^{}_{\mathbf{k}}) \,
\delta( B^{\dagger}_{\mathbf{-q}, \lambda}+ B_{\mathbf q, \lambda} ) + \textrm{H.c.} \Big]  \, . \nonumber \\
&
\end{align}
The parameters of $\mathcal H_{0,\lambda}$ depend on the cutoff $\lambda$.
Also the phonon energy has acquired an additional $\bf q$-dependence. 
Moreover,
we have introduced  $\lambda$-dependent phonon operators 
 \begin{eqnarray}
 \label{A4}
B^\dag_{\bf q,\lambda} = b_{\bf q}^\dag + \frac{\sqrt N h_{ \lambda}}{\omega_{\bf Q,\lambda}} 
 \delta_{\bf q, \bf Q} \, ,
 \end{eqnarray}
in a slight generalization of the former definition \eqref{npo}. Finally, the quantity $\mathbf P_\lambda$ 
in Eq.~\eqref{A3} is a generalized projector, which projects on all transitions (with respect to $\mathcal H_{0,\lambda}$) 
with energies smaller than $\lambda$.
Note that the coupling strength $g$ of $\mathcal H_{1,\lambda}$ remains
$\lambda$-independent as a consequence of the present restriction to
renormalization contributions up to order $g^2$.  
 
Next $\mathbf P_\lambda$ has to be applied 
to the operators $ \delta(c^\dagger_{\mathbf{k+q}}f^{}_{\mathbf{k}}) \,
\delta( B^{\dagger}_{\mathbf{-q}, \lambda}+ B_{\mathbf q, \lambda}) $ in $\mathcal H_{1,\lambda}$, 
which requires the decomposition of the operators in the squared brackets 
into dynamical eigenmodes of $\mathcal H_{0,\lambda}$. 
One may show  that one can use for $\mathcal H_{1,\lambda}$ 
 \begin{eqnarray}
\label{A5}
\mathcal H_{1,\lambda} &=&
  \frac{g}{\sqrt{N}}\sum_{\mathbf{k} \mathbf q  } 
  \Big[ \Theta^{+}_{\bf k \bf q, \lambda} \big( \delta (c^\dagger_{\mathbf{k+q}}f^{}_{\mathbf{k}}) \,
 \delta B^{\dagger}_{\mathbf{-q}, \lambda} +  {\rm H.c.}\big) \nonumber \\
 && \qquad  +
 \Theta^{-}_{\bf k \bf q, \lambda} \big(\delta(c^\dagger_{\mathbf{k+q}}f^{}_{\mathbf{k}}) \,
 \delta B_{\mathbf q, \lambda})  + \textrm{H.c.}\big) \Big] 
 \end{eqnarray}
 as long as one is only interested in renormalization equations up to linear order in the order parameters.  
In Eq.~\eqref{A5} we have introduced two $\Theta$-functions
 \begin{eqnarray}
\label{A6}
 \Theta^{\pm}_{\bf k \bf q, \lambda} &=& 
 \Theta(\lambda -|\varepsilon^c_{\bf k+ \bf q, \lambda} - \varepsilon^f_{\bf k, \lambda} 
 \pm \omega_{\mp\bf q, \lambda}|)\, ,
 \end{eqnarray}
which restrict transitions to excitation energies smaller than $\lambda$.  

One can also construct the generator  $X_{\lambda, \Delta \lambda}$ of the unitary
transformation \eqref{A1} for the transformation from cutoff $\lambda$ to $\lambda- \Delta \lambda$.
According to Ref.~\onlinecite{BHS02} the lowest order
for $X_{\lambda, \Delta \lambda}$ is given by
\begin{eqnarray}
\label{A7}
X_{\lambda, \Delta \lambda} = \frac{1}{ \mathbf L_{0,\lambda}} \mathbf Q_{\lambda- \Delta \lambda} \mathcal H_{1,\lambda}
\, .
\end{eqnarray}
Here $ \mathbf L_{0,\lambda}$ is the Liouville operator of the unperturbed Hamiltonian $\mathcal H_{0,\lambda}$, which is 
defined by $ \mathbf L_{0,\lambda} \mathcal A = [\mathcal{H}_{0, \lambda}, \mathcal A]$ for any operator quantity 
$\mathcal A$, and $\mathbf Q_{\lambda - \Delta \lambda} =1 - \mathbf P_{\lambda - \Delta \lambda}$ is the complement 
projector to $\mathbf P_{\lambda - \Delta \lambda}$. It projects on all transition operators with excitation energies 
larger than $\lambda -\Delta \lambda$.  
With Eqs.~\eqref{A5} and \eqref{A2} one finds
 \begin{align}
  \label{A8} 
X_{\lambda, \Delta \lambda} &= \frac{g}{\sqrt N} \sum_{\bf k \bf q } \Big[
A^+_{\bf k \bf q}(\lambda, \Delta \lambda) \, 
 \big(\delta(c^\dagger_{\mathbf{k+q}}f^{}_{\mathbf{k}}) \delta B^{\dagger}_{\mathbf{-q},\lambda}  - \textrm{H.c.}\big) \nonumber \\
 &+ A^-_{\bf k \bf q}(\lambda, \Delta \lambda) \, 
\big(\delta(c^\dagger_{\mathbf{k+q}}f^{}_{\mathbf{k}})
\delta B_{\mathbf q, \lambda}  - \textrm{H.c.} \big)
\Big] \, ,
 \end{align}
where the prefactors are given by 
\begin{eqnarray}
  \label{A9} 
&& A^{\pm}_{\bf k \bf q}(\lambda, \Delta \lambda) =
\frac{\Theta^\pm _{\bf k \bf q, \lambda} \big( 1- \Theta^\pm_{\bf k \bf q, \lambda- \Delta \lambda}\big)}
{\varepsilon^c_{\bf k + \bf q, \lambda} - \varepsilon^f_{\bf k, \lambda} \pm \omega_{\mp\bf q, \lambda}}\, .
 \end{eqnarray}
Here the products of the two $\Theta$-functions in $A^\pm_{\bf k \bf q}(\lambda, \Delta \lambda)$ 
assure that only excitations between $\lambda$
and $\lambda -\Delta \lambda$ are eliminated by the unitary transformation \eqref{A1}.  
Also note that the Liouville operator $\mathbf L_{0,\lambda}$ in $X_{\lambda, \Delta \lambda}$ (and the projector 
$\mathbf P_\lambda$ in $\mathcal H_{1,\lambda}$) in principle should have been defined with respect to the full 
unperturbed Hamiltonian 
$\mathcal H_{0,\lambda}$ of Eq.~\eqref{A2} and not by leaving out the last term $\propto \Delta_{\lambda}$. 
However, inclusion of this term would only give small higher-order corrections to $\Delta_\lambda$. 


\subsection{Renormalization equations}
\label{A:2}

The $\lambda$-dependence of the parameters of $\mathcal H_\lambda$ are found from 
transformation \eqref{A1}.  
For small enough width $\Delta \lambda$ of the transformation steps, the expansion of  \eqref{A1} in $g$ can be limited to ${\cal O}(g^2)$ terms. 
One obtains
\begin{eqnarray}
  \label{A10} 
\mathcal H_{\lambda -\Delta \lambda} &=&  \mathcal H_{0,\lambda} + \mathbf P_{\lambda - \Delta \lambda} 
\mathcal H_{1,\lambda}
+ [X_{\lambda, \Delta \lambda} , \mathcal H_{1,\lambda}]  \nonumber \\
&-& \frac{1}{2} [X_{\lambda, \Delta \lambda} , \mathbf Q_{\lambda- \Delta \lambda} \mathcal H_{1, \lambda}] + \cdots \, ,
   \end{eqnarray}
where Eq.~\eqref{A7} has been used. Renormalization contributions to $\mathcal H_{\lambda - \Delta \lambda}$ 
arise from the last two commutators which have to be evaluated explicitly. The result
must be compared with the generic form \eqref{A2}, \eqref{A3} of $\mathcal H_\lambda$ (with $\lambda$ 
replaced by $\lambda - \Delta \lambda$) when it is written in terms of the original  $\lambda$-independent variables 
$c^\dag_{\bf k}$, $f^\dag_{\bf k}$, and $b^{\dag}_{\bf q}$.  
This leads to the 
 following renormalization equations for the parameters of $\mathcal H_{0,\lambda}$:
\begin{align}
\label{A11}
&\varepsilon^c_{\mathbf{k},\lambda-\Delta\lambda}-\varepsilon^c_{\mathbf{k},\lambda}= \nonumber \\
& \quad \frac{2g^2}{N}\sum_{\mathbf{q}} \Big(A^+_{\mathbf{k-q,q}}(\lambda,\Delta\lambda)\, 
(n^B_{\mathbf{-q}}+n^f_{\mathbf{k-q}})  \nonumber \\
&\quad + A^-_{\mathbf{k-q,q}}(\lambda,\Delta\lambda)\, (1+ n^B_{\mathbf{q}}- n^f_{\mathbf{k-q}}) 
\Big) \,,\\[0.2cm]
%
&\varepsilon^f_{\mathbf{k},\lambda-\Delta\lambda}- \varepsilon^f_{\mathbf{k},\lambda} = \nonumber \\
& \quad -\frac{2g^2}{N}\sum_{\mathbf{q}}\Big(A^+_{\mathbf{kq}}(\lambda,\Delta\lambda)(1-n^c_{\mathbf{k+q}}
+n^B_{\mathbf{-q}}) 
 \nonumber\\
& \quad + A^-_{\mathbf{kq}}(\lambda,\Delta\lambda)(n^c_{\mathbf{k+q}}+n^B_{\mathbf{-q}}) 
\Big)\,,  
\label{A12}
\end{align}
and 
\begin{align}
\label{A13}
&\omega_{\mathbf{q},\lambda-\Delta\lambda}- \omega_{\mathbf{q},\lambda}=    
- \frac{2g^2}{N}\sum_{\mathbf{k}}
\big[A^+_{\mathbf{k,-q}}(\lambda, \Delta \lambda) (n^f_{\mathbf{k}}- n^c_{\mathbf{k-q}}) \nonumber \\
&\qquad+ A^-_{\mathbf{k,q}}(\lambda, \Delta \lambda) (n^f_{\mathbf{k}}- n^c_{\mathbf{k+q}}) \big] 
\, , \\[0.2cm]
&h_{\lambda - \Delta \lambda} - h_{\lambda} = - \frac{g^2}{N\sqrt{N}}  \sum_{\mathbf{k}} 
\big[A^+_{\mathbf{k,-Q}}(\lambda, \Delta \lambda) (n^c_{\mathbf{k-Q}}- n^f_{\mathbf{k}}) \nonumber \\
&\qquad+ A^-_{\mathbf{k,Q}}(\lambda, \Delta \lambda) (n^c_{\mathbf{k+Q}}- n^f_{\mathbf{k}}) \big]
\big(\langle b_{\bf Q} \rangle  + \langle b^\dag_{\bf Q} \rangle \big)\,, 
\label{A14} 
\\[0.2cm]
& \Delta_{\lambda-\Delta\lambda}-\Delta_{\lambda} \simeq 0  \, .  
\label{A15}
\end{align}
The quantities $n^c_{\bf k}$, $  n^f_{\bf k}$ and $n^B_{\bf q}$ 
are occupation numbers for electrons and phonons,
 \begin{eqnarray}
  \label{A16}
&& n^c_{\bf k} = \langle c^\dag_{\bf k} c_{\bf k}\rangle \, , \quad 
    n^f_{\bf k} = \langle f^\dag_{\bf k} f_{\bf k}\rangle  \\
    && \nonumber \\
     \label{A17}
 && 
 n^B_{\bf q} = \langle \delta B_{\bf q,\lambda}^\dag \delta B_{\bf q, \lambda}\rangle =
 \langle \delta b_{\bf q}^\dag \delta b_{\bf q}\rangle \, ,
\end{eqnarray} 
and have to be evaluated separately. Note that also  $ n^B_{\bf q}$ is $\lambda$-independent, 
which was already used in the renormalization equations.
For the numerical solution in Section~\ref{S:V} the initial parameter values at 
$\lambda = \Lambda$ are needed, which are 
those of the original Hamiltonian $\mathcal{H}$  
\begin{eqnarray}
  \label{A18} 
  {\varepsilon}^f_{\mathbf{k}, \Lambda} = {\varepsilon}^f_{\mathbf{k}} \,, \quad
 {\varepsilon}^c_{\mathbf{k}, \Lambda} = {\varepsilon}^c_{\mathbf{k}} \,,  \quad
  \omega_{{\bf q}, \Lambda} =  \omega_{0}\,, 
  \end{eqnarray}
and 
\begin{eqnarray}
\label{A19}
&& h_{\Lambda} = h = 0^+ \, , \quad 
  \Delta_\Lambda = \Delta = 0^+
  \, . 
  \end{eqnarray}
Suppose the expectation values in \eqref{A11}- \eqref{A15} are already known, the renormalization equations 
can be integrated between $\lambda = \Lambda$ and $\lambda =0$. 
In this way, we obtain the fully renormalized Hamiltonian $\tilde{\mathcal{H}}:= \mathcal H_{\lambda=0} =  
\mathcal H_{0, \lambda=0}$, as was already stated in Eq.~\eqref{10a}
\begin{eqnarray}
\label{A20}
&& \tilde{\mathcal{H}} =
\sum_{\mathbf{k}}\tilde{\varepsilon}^f_{\mathbf{k}}f^\dagger_{\mathbf{k}}f^{}_{\mathbf{k}}
+\sum_{\mathbf{k}}\tilde{\varepsilon}^c_{\mathbf{k}}c^\dagger_{\mathbf{k}}c^{}_{\mathbf{k}}
+ \sum_{\mathbf{q}}\tilde{\omega}_\mathbf{q}
b^\dag_{\mathbf{q}} b^{}_{\mathbf{q}}
\\
&& \quad +  \tilde{\Delta}\sum_{\mathbf{k}}(c^\dagger_{\mathbf{k+Q}}f^{}_{\mathbf{k}}+\textrm{H.c.}) 
+ {\sqrt N}\, \tilde h (b^\dag_{-\bf Q} + b_{-\bf Q} )  \, . \nonumber 
\end{eqnarray}
The  tilde symbols denote the fully renormalized quantities at  $\lambda =0$.  All  excitations from
$\mathcal H_{1,\lambda}$ with non-zero energies have been eliminated.
They give rises to the renormalization of $\mathcal H_{0,\lambda}$. Note that the order 
parameter $\Delta$ remains unrenormalized, i.e.~$\tilde \Delta = \Delta$, due to renormalization equation
\eqref{A15}.

Finally, Eq.~\eqref{A20} can be expressed in terms of the renormalized boson operators $\tilde B^\dag_{\bf q} 
= b_{\bf q}^\dag + ({\sqrt N \tilde h}/{\tilde \omega_{\bf Q}})  \delta_{\bf q, \bf Q}$, which gives 
\begin{eqnarray}
\label{A21}
&& \tilde{\mathcal{H}} =
\sum_{\mathbf{k}}\tilde{\varepsilon}^f_{\mathbf{k}}f^\dagger_{\mathbf{k}}f^{}_{\mathbf{k}}
+\sum_{\mathbf{k}}\tilde{\varepsilon}^c_{\mathbf{k}}c^\dagger_{\mathbf{k}}c^{}_{\mathbf{k}}
+ \sum_{\mathbf{q}}\tilde{\omega}_\mathbf{q}
\tilde B^\dag_{\mathbf{q}} \tilde B^{}_{\mathbf{q}}
 \nonumber \\
&& \quad +  \tilde{\Delta}\sum_{\mathbf{k}}(c^\dagger_{\mathbf{k+Q}}f^{}_{\mathbf{k}}+\textrm{H.c.})  \,. 
\end{eqnarray}
As in the unrenormalized case, the electronic part of $\tilde{\mathcal H}$ will be diagonalized by a 
Bogoliubov transformation, which gives 
\begin{eqnarray}
\label{A22}
\tilde{\mathcal{H}}&=&\sum_{\mathbf{k}} \tilde E^{(1)}_{\mathbf{k}}{C}^{\dagger}_{1, \mathbf{k}}{C}_{1, \mathbf{k}}
+\sum_{\mathbf{k}}\tilde E^{(2)}_{\mathbf{k}}{C}^{\dagger}_{2, \mathbf{k}}{C}_{2, \mathbf{k}} \nonumber \\
&+& \sum_{\bf q} \tilde{\omega}_{\bf q} \tilde{B}^\dag_{\bf q} \tilde{B}_{\bf q} + {\rm const.} \, .
\end{eqnarray}
In result \eqref{A22} the electronic quasiparticle energies $\tilde E^{(1,2)}_{\bf k}$ and 
also the quasiparticle modes 
$C^{(\dag)}_{1,\bf k}$, $C^{(\dag)}_{2,\bf k}$
are  now renormalized quantities. They are defined by the former equations \eqref{9d}- \eqref{9g}
when the unrenormalized energies $\varepsilon_{\bf k}^c$,  $\varepsilon_{\bf k}^f$ are replaced 
by the renormalized  energies $\tilde \varepsilon_{\bf k}^c$,  $\tilde\varepsilon_{\bf k}^f$. 
Note that the quadratic form of Eq.~\eqref{A22} allows to compute any expectation value formed with 
$\tilde{\mathcal H}$.

\subsection{Expectation values}
\label{A:3}

Next, expectation values $\langle  \mathcal A \rangle$, formed with the full $\mathcal H$, have to be evaluated in the framework of the PRM.
As already stated in Sec.~\ref{S:IV}, they can be found by 
exploiting the unitary invariance of operator expressions below a trace. 
Employing the same unitary transformation to $\mathcal A$ as before for 
the Hamiltonian,\cite{BHS02} one finds 
$\langle {\mathcal A} \rangle = \langle \mathcal{A}(\lambda)\rangle_{\mathcal H_\lambda}  =
 \langle \tilde{\mathcal A} \rangle_{\tilde{\mathcal H}}$. 
Here $\mathcal A(\lambda)= e^{X_\lambda} {\mathcal A}e^{-X_\lambda}$ and  $\tilde{\mathcal A} = \mathcal A(\lambda =0)$.
$X_\lambda$ is the generator for the unitary transformation between cutoff $\Lambda$ and $\lambda$.
To find the expectation values of Eqs.~\eqref{A16}, \eqref{A17} one best starts from an
{\it ansatz} for the single fermion operators 
$c^\dagger_{\mathbf{k}}(\lambda)=e^{X_\lambda}c^\dagger_{\mathbf{k}}e^{-X_\lambda}$,
$f^\dagger_{\mathbf{k}}(\lambda)=e^{X_\lambda}f^\dagger_{\mathbf{k}}e^{-X_\lambda}$, and the phonon operator
$b^\dagger_{\mathbf{q}}(\lambda)=e^{X_\lambda}b^\dagger_{\mathbf{q}}e^{-X_\lambda}$,
at cutoff $\lambda$. 
 In  second order in the electron-phonon interaction they are chosen as 
\begin{align}
\label{A23}
c^\dagger_{\mathbf{k}}(\lambda)&=x_{\mathbf{k},\lambda}c^\dagger_{\mathbf{k}}+
\frac{1}{\sqrt{N}}\sum_{\mathbf{q}}t^+_{\mathbf{k-q,q},\lambda}f^\dagger_{\mathbf{k-q}}\delta(B_{\mathbf{-q},\lambda})\nonumber\\
&+\frac{1}{\sqrt{N}}\sum_{\mathbf{q}}t^-_{\mathbf{k-q,q},\lambda}f^\dagger_{\mathbf{k-q}}\delta(B^\dagger_{\mathbf{q},\lambda})\,,\\
f^\dagger_{\mathbf{k}}(\lambda)&=y_{\mathbf{k},\lambda}f^\dagger_{\mathbf{k}}
+\frac{1}{\sqrt{N}}\sum_{\mathbf{q}}u^+_{\mathbf{kq},\lambda}c^\dagger_{\mathbf{k+q}}\delta(B^\dagger_{\mathbf{-q},\lambda})\nonumber\\
&+\frac{1}{\sqrt{N}}\sum_{\mathbf{q}}u^-_{\mathbf{kq},\lambda}c^\dagger_{\mathbf{k+q}}\delta(B^{}_{\mathbf{q},\lambda})\,,
\label{A24}\\
b^\dagger_{\mathbf{q}}(\lambda)&=z_{\mathbf{q},\lambda}b^\dagger_{\mathbf{q}}
+\frac{1}{\sqrt{N}}\sum_{\mathbf{k}}v^+_{\mathbf{k,-q},\lambda}\delta(f^\dagger_{\mathbf{k}}c^{}_{\mathbf{k-q}})\nonumber\\
&+\frac{1}{\sqrt{N}}\sum_{\mathbf{k}}v^-_{\mathbf{kq},\lambda}\delta(c^\dagger_{\mathbf{k+q}}f^{}_{\mathbf{k}})\,.
\label{25}
\end{align}
In analogy to the renormalization equations for the parameters of $\mathcal H_\lambda$, 
one first derives the following set of renormalization equations for the coefficients 
$t^\pm_{\mathbf{k-q,q},\lambda}$, $u^\pm_{\mathbf{kq},\lambda}$, and $v^\pm_{{\mathbf{k},\mp\mathbf{q}},\lambda}$:
\begin{eqnarray}
\label{A26}
t^{\pm}_{\mathbf{k-q,q},\lambda-\Delta\lambda}&=&t^\pm_{\mathbf{k-q,q},\lambda}-gx^{}_{\mathbf{k},\lambda}A^\pm_{\mathbf{k-q,q}}
(\lambda,\Delta\lambda)\,,\\[0.2cm]
u^\pm_{\mathbf{kq},\lambda-\Delta\lambda}&=&u^\pm_{\mathbf{kq},\lambda}+gy_{\mathbf{k},\lambda}A^\pm_{\mathbf{kq}}
(\lambda,\Delta\lambda)\,,
\label{A27}\\[0.2cm]
v^\pm_{{\mathbf{k},\mp\mathbf{q}},\lambda-\Delta\lambda}&=&v^\pm_{{\mathbf{k},\mp\mathbf{q}},\lambda}
-gz_{\mathbf{q},\lambda}A^\pm_{\mathbf{k},\mp\mathbf{q}}(\lambda,\Delta\lambda)\,.\label{A28}
\end{eqnarray}
Using the anticommutation relations for fermion operators and the commutation relations for boson operators  
(as for instance $[c^\dagger_{\mathbf{k}}(\lambda),c_{\mathbf{k}}(\lambda)]_+=1$, valid for any $\lambda$) 
one arrives at 
\begin{align}
\label{A29}
|x_{\mathbf{k},\lambda}|^2=&1-\frac{1}{N}\sum_{\mathbf{q}}\Big[|t^+_{\mathbf{k-q,q},\lambda}|^2
(n^B_{\mathbf{-q},\lambda}+n^f_{\mathbf{k-q}})\nonumber\\
&\quad\quad\quad+|t^-_{\mathbf{k-q,q},\lambda}|^2
(1+n^B_{\mathbf{q},\lambda}-n^f_{\mathbf{k-q}})\Big]\,,\\[0.2cm]
\label{A30}
|y_{\mathbf{k},\lambda}|^2=&1-\frac{1}{N}\sum_{\mathbf{q}}\Big[|u^+_{\mathbf{kq},\lambda}|^2
(n^B_{\mathbf{-q},\lambda}+1-n^c_{\mathbf{k+q}})\nonumber\\
&\quad\quad\quad+|u^-_{\mathbf{kq},\lambda}|^2
(n^B_{\mathbf{q},\lambda}+n^c_{\mathbf{k+q}})\Big]\,,\\[0.2cm]
\label{A31}
|z_{\mathbf{q},\lambda}|^2=&1-\frac{1}{N}\sum_{\mathbf{k}}\Big[|v^+_{\mathbf{k,-q},\lambda}|^2
(n^c_{\mathbf{k-q}}-n^f_{\mathbf{k}})\nonumber\\
&\quad\quad\quad+|v^-_{\mathbf{kq},\lambda}|^2
(n^f_{\mathbf{k}}-n^c_{\mathbf{k+q}})\Big]\,.
\end{align}
Note that Eqs.~\eqref{A26}-\eqref{A28} together with the new set  \eqref{A29}-\eqref{A31}, 
taken at $\lambda \rightarrow \lambda -\Delta \lambda$, represents
a complete set of renormalization  equations for all $\lambda$-dependent coefficients in Eqs.~\eqref{A23}-\eqref{25}. 
They combine the parameter values at $\lambda$ with those at $\lambda - \Delta \lambda$. 
By integrating the full set
 between $\lambda=\Lambda$, with initial parameter values
\begin{align}
\label{A32}
\{x_{\mathbf{k}\Lambda},y_{\mathbf{k}\Lambda},z_{\mathbf{k}\Lambda}\}=1, \quad
\{t^\pm_{\mathbf{kq},\Lambda},u^\pm_{\mathbf{kq},\Lambda},v^\pm_{\mathbf{kq},\Lambda}\}=0\, ,
\end{align}
and $\lambda=0$,  one is led to the fully renormalized one-particle operators
\begin{align}
\label{A33}
\tilde{c}^\dagger_{\mathbf{k}}=\tilde{x}_{\mathbf{k}}c^\dagger_{\mathbf{k}}
&+\frac{1}{\sqrt{N}}\sum_{\mathbf{q}}\tilde{t}^+_{\mathbf{k-q,q}}f^\dagger_{\mathbf{k-q}}\delta(\tilde{B}^{}_{\mathbf{-q}})\nonumber\\
&+\frac{1}{\sqrt{N}}\sum_{\mathbf{q}}\tilde{t}^-_{\mathbf{k-q,q}}f^\dagger_{\mathbf{k-q}}\delta(\tilde{B}^\dagger_{\mathbf{q}})\,,\\
\tilde{f}^\dagger_{\mathbf{k}}=\tilde{y}_{\mathbf{k}}f^\dagger_{\mathbf{k}}
&+\frac{1}{\sqrt{N}}\sum_{\mathbf{q}}\tilde{u}^+_{\mathbf{kq}}c^\dagger_{\mathbf{k+q}}\delta(\tilde{B}^\dagger_{\mathbf{-q}})\nonumber\\
&+\frac{1}{\sqrt{N}}\sum_{\mathbf{q}}\tilde{u}^-_{\mathbf{kq}}c^\dagger_{\mathbf{k+q}}\delta(\tilde B^{}_{\mathbf{q}})\,,
\label{A34}\\
\tilde{b}^\dagger_{\mathbf{q}}=\tilde{z}_{\mathbf{q}}b^\dagger_{\mathbf{q}}
&+\frac{1}{\sqrt{N}}\sum_{\mathbf{k}}\tilde{v}^+_{\mathbf{k,-q}}\delta(f^\dagger_{\mathbf{k}}c^{}_{\mathbf{k-q}})\nonumber\\
&+\frac{1}{\sqrt{N}}\sum_{\mathbf{k}}\tilde{v}^-_{\mathbf{kq}}\delta(c^\dagger_{\mathbf{k+q}}f^{}_{\mathbf{k}}) \, .
\label{A35}
\end{align}
As before, the tilde symbols denote fully renormalized quantities. 
With Eqs.~\eqref{A33}-\eqref{A35} the expectation values \eqref{A16}, \eqref{A17} can be evaluated.  
The expectation values for fermion operators read up to order ${\cal O}(g_{\mathbf{k}}^2)$: 
\begin{align}
\label{A36}
&n^c_{\mathbf{k}}=|\tilde{x}_{\mathbf{k}}|^2\tilde{n}^c_{\mathbf{k}}+\frac{1}{N}\sum_{\mathbf{q}}
\Big[|\tilde{t}^+_{\mathbf{k-q,q}}|^2\tilde{n}^f_{\mathbf{k-q}}(1+\tilde{n}^B_{\mathbf{-q}})\nonumber\\
&\quad\quad\quad\quad\quad\quad\quad\quad\quad+|\tilde{t}^-_{\mathbf{k-q,q}}|^2\tilde{n}^f_{\mathbf{k-q}}\tilde{n}^B_{\mathbf{q}}\Big]\,,\\
\label{A37}
&n^f_{\mathbf{k}}=|\tilde{y}_{\mathbf{k}}|^2\tilde{n}^f_{\mathbf{k}}+\frac{1}{N}\sum_{\mathbf{q}}
\Big[|\tilde{u}^+_{\mathbf{kq}}|^2\tilde{n}^c_{\mathbf{k+q}}\tilde{n}^B_{\mathbf{-q}}\nonumber\\
&\quad\quad\quad\quad\quad\quad\quad\quad\quad+|\tilde{u}^-_{\mathbf{kq}}|^2\tilde{n}^c_{\mathbf{k+q}}(1+\tilde{n}^B_{\mathbf{q}})\Big]\,,\\
&d_{\bf k}=\tilde{x}_{\mathbf{k+Q}}
\tilde{y}_{\mathbf{k}}\langle c^\dagger_{\mathbf{k+Q}}f^{}_{\mathbf{k}}\rangle_{\tilde{\mathcal{H}}}
\nonumber\\[0.1cm]
&+\frac{1}{N}\sum_{\mathbf{q}}
\Big[\tilde{t}^+_{\mathbf{k+Q-q},\mathbf{q}}\tilde{u}^-_{\mathbf{k},-\mathbf{q}}
\langle f^\dagger_{\mathbf{k+Q-q}}c^{}_{\mathbf{k-q}}\rangle_{\tilde{\mathcal{H}}}(1-\tilde{n}^B_{\mathbf{-q}})\nonumber\\
&\hspace*{1.3cm}+\tilde{t}^-_{\mathbf{k+Q-q},\mathbf{q}}\tilde{u}^+_{\mathbf{k,-q}}
\langle f^\dagger_{\mathbf{k+Q-q}}c^{}_{\mathbf{k-q}}\rangle_{\tilde{\mathcal{H}}}\tilde{n}^B_{\mathbf{q}}\Big]
\label{A38}
 \, .
\end{align}
Here $d_{\bf k}= \langle c^\dag_{\bf k + \bf Q} f_{\bf k}\rangle$[Eq.~\eqref{3}] is an additional quantity, which 
acts as an excitonic order parameter. The expectation values on the right-hand sides of Eqs.~\eqref{A39}-\eqref{A41}
are formed with $\tilde{\mathcal H}$ and can be  evaluated, i.e.
\begin{eqnarray}
\label{A39}
&&\tilde{n}^c_{\mathbf{k+Q}}=\langle c^\dagger_\mathbf{k+Q}c_\mathbf{k+Q}\rangle_{\tilde{\mathcal{H}}} = 
\xi^2_{\mathbf k}f^F(\tilde E^1_{\mathbf k})+\eta^2_{\mathbf k}f^F(\tilde E^2_{\mathbf k})\,,\\[0.2cm]
\label{A40}
&&\tilde{n}^f_{\mathbf{k}}=\langle f^\dagger_{\mathbf k}f_\mathbf{k}\rangle_{\tilde{\mathcal{H}}}=
\eta^2_{\mathbf k}f^F(\tilde E^1_{\mathbf k})+\xi^2_{\mathbf k}f^F(\tilde E^2_{\mathbf k}) \,, \\[0.2cm]
&& \langle c^\dagger_{\mathbf{k+Q}}f^{}_{\mathbf{k}}\rangle_{\tilde{\mathcal{H}}} 
=-[f^F(\tilde E^1_{\mathbf k})-f^F(\tilde E^2_{\mathbf k})]\textrm{sgn}(\tilde{\varepsilon}^f_{\mathbf k}
-\tilde{\varepsilon}^c_\mathbf{k+Q})\frac{\tilde{\Delta}}{W_{\mathbf k}}. 
\label{A41}\nonumber\\
\end{eqnarray}
Here the prefactors $\xi_{\bf k}$ and $\eta_{\bf k}$ are the coefficients from the Bogoliubov transformation, used in
Eq.~\eqref{A22}. As mentioned, they are defined by Eqs.~\eqref{9f}, \eqref{9fa}, when the unrenormalized one-particle energies are replaced by the renormalized ones.  

The bosonic expectation value \eqref{A17} is given by
\begin{align}
\label{A42}
&n^B_{\bf q}  = 
\langle \delta b^\dag_{\bf q} \delta b_{\bf q}\rangle =
\langle b^\dag_{\bf q} b_{\bf q}\rangle - 
\langle b^\dag_{\bf q} \rangle  \langle b^\dag_{\bf q} \rangle \delta_{\bf q= \bf Q}\, ,
\end{align}
 where
\begin{align}
\label{A43}
&\langle b^\dag_{\bf q} b_{\bf q}\rangle=|\tilde{z}_{\mathbf{q}}|^2\tilde{n}^b_{\mathbf{q}}+\frac{1}{N}\sum_{\mathbf{k}}
\Big[|\tilde{v}^+_{\mathbf{k,-q}}|^2\tilde{n}^f_{\mathbf{k}}(1-\tilde{n}^c_{\mathbf{k-q}})\nonumber\\
&\hspace*{3cm}+|\tilde{v}^-_{\mathbf{kq}}|^2\tilde{n}^c_{\mathbf{k+q}}(1-\tilde{n}^f_{\mathbf{k}})\Big] \,,\\
& \langle b^\dag_{\bf q} \rangle \simeq\tilde z_{\bf q}    \langle b^\dag_{\bf q} \rangle_{\tilde{\mathcal H}} \, .
\label{A44}
\end{align}
We emphasize that in $\langle b^\dag_{\bf q}\rangle$ smaller 
contributions from  \eqref{A35} have been neglected. Using Eq.~\eqref{A21} the 
expectation values 
 $\tilde{n}^b_{\mathbf{q}}= \langle b^\dag_{\bf q} b_{\bf q}  \rangle_{\tilde{\mathcal H}}$ and 
 $ \langle b^\dag_{\bf q} \rangle_{\tilde{\mathcal H}}$ on the right hand sides become  
 \begin{align}
\label{A45}
\tilde{n}^b_{\mathbf{q}}&=\langle B^\dagger_\mathbf{q} B^{}_\mathbf{q}\rangle_{\tilde{\mathcal{H}}}-\frac{\sqrt{N}\tilde{h}}{\tilde{\omega}_\mathbf{q}}
\langle B^\dagger_\mathbf{q} + B^{}_\mathbf{q}\rangle_{\tilde{\mathcal{H}}}\delta_{\mathbf{q,Q}}
+\frac{N\tilde{h}^2}{\tilde{\omega}^2_\mathbf{q}}\delta_{\mathbf{q,Q}}\nonumber\\
&=f^B(\tilde{\omega}_\mathbf{q})+\frac{N\tilde{h}^2}{\tilde{\omega}^2_\mathbf{q}}\delta_{\mathbf{q,Q}}
\end{align}
and 
\begin{equation}
\label{A46}
 \langle b^\dag_{\bf q} \rangle_{\tilde{\mathcal H}} = \Big[\langle B^\dag_{\bf q} \rangle_{\tilde{\mathcal H}} 
 - \frac{\sqrt N  \, \tilde h}{\tilde \omega_{\bf q}} \Big] \delta_{\bf q, \bf Q} =  
 - \frac{\sqrt N \, \tilde h}{\tilde \omega_{\bf q}}  \delta_{\bf q, \bf Q} \,.
\end{equation} 
Here we have used $\langle B^\dag_{\bf q} \rangle_{\tilde{\mathcal H}} =0$.   $f^B(\tilde{\omega}_{\mathbf k})$ is the bosonic distribution function which becomes one at zero temperature. Inserting Eqs.~\eqref{A45}, \eqref{A46} into \eqref{A43}, \eqref{A44}  one finally arrives at 
\begin{align}
\label{A47}
& n^B_{\bf q}  =|\tilde{z}_{\mathbf{q}}|^2 f^B(\tilde{\omega}_\mathbf{q})
+\frac{1}{N}\sum_{\mathbf{k}}
\Big[|\tilde{v}^+_{\mathbf{k,-q}}|^2\tilde{n}^f_{\mathbf{k}}(1-\tilde{n}^c_{\mathbf{k-q}})\nonumber\\
&\hspace*{2.8cm} +|\tilde{v}^-_{\mathbf{kq}}|^2\tilde{n}^c_{\mathbf{k+q}}(1-\tilde{n}^f_{\mathbf{k}})\Big]  \,.
 \end{align}
Note that the electronic order parameter $d_{\bf k}$ and the phononic 
order parameter $\Delta$ are intimately related. Due to \eqref{A38} and  \eqref{A41}, 
$d_{\bf k}$ is proportional to $\tilde{\Delta}= \Delta$, so that both order parameters are 
mutually dependent. Finally, as a side remark, note that the lattice displacement in the EI state is given by 
\begin{equation}
\label{A48}
x^{}_\mathbf{Q}=\frac{1}{\sqrt{N}} \frac{\tilde{z}_\mathbf{Q}}{\sqrt{2\tilde{\omega}_\mathbf{Q}}}
\langle b^{\dagger}_\mathbf{-Q}+ b^{}_\mathbf{Q}\rangle_{\tilde{\mathcal{H}}}=-\sqrt{\frac{2}{\tilde{\omega}_\mathbf{Q}}} \, 
\frac{\tilde{h}\, \tilde{z}_\mathbf{Q}}{\tilde{\omega}_\mathbf{Q}}\,,
\end{equation}
as follows from Eqs.~\eqref{A44} and \eqref{A46}. 

\subsection{Spectral functions}
\label{A:4}

Let  us first evaluate the electronic one-particle spectral functions. 
Here, the $c$-electron spectral function $A^c_{\bf k}(\omega)$ was defined before in Eq.~\eqref{9i}, 
 \begin{eqnarray*}
 A^c_{\bf k}(\omega) &=& \frac{1}{2\pi} \int_{-\infty}^\infty \langle [c_{\bf k \sigma}(t), c^\dag_{\bf k \sigma}]_+\rangle
 e^{i\omega t} dt \, , 
\end{eqnarray*}
where the expectation value is formed with $\mathcal H$. Using the unitary invariance of 
operator expressions under a trace one arrives at the former expression \eqref{10b}.  By use of Eq.~\eqref{A33}
and Eq.~\eqref{A22} the following result for $ A^c_{\bf k}(\omega)$ is found
\begin{align}
\label{A49}
A^c_{\mathbf k}(\omega)=&|\tilde{x}_{\mathbf{k}}|^2[\xi^2_{\mathbf{k-Q}}\delta(\omega-\tilde E^1_{\mathbf{k-Q}})
+\eta^2_{\mathbf{k-Q}}\delta(\omega-\tilde E^2_{\mathbf{k-Q}})]\nonumber\\[0.1cm]
&\hspace*{-0.8cm}+\frac{1}{N}\sum_{\mathbf{q}}\Big[|\tilde{t}^+_{\mathbf{k-q,q}}|^2(\tilde{n}^B_{\mathbf{-q}}+\tilde{n}^f_{\mathbf{k-q}})
\delta(\omega-\tilde{\varepsilon}^f_{\mathbf{k-q}}+\tilde{\omega}^{}_{\mathbf{-q}})\nonumber\\
&\hspace*{-0.8cm}+|\tilde{t}^-_{\mathbf{k-q,q}}|^2(1+\tilde{n}^B_{\mathbf{q}} - \tilde{n}^f_{\mathbf{k-q}})
\delta(\omega-\tilde{\varepsilon}^f_{\mathbf{k-q}}-\tilde{\omega}^{}_{\mathbf{q}})\Big]\,.
\end{align}
Similarly, for the $f$-electron spectral function one finds
\begin{eqnarray}
\label{A50}
&& A^f_{{\mathbf k}}(\omega)=|\tilde{y}_{\mathbf{k}}|^2[\eta^2_{\mathbf k}\delta(\omega-\tilde E^1_{\mathbf k})
+\xi^2_{\mathbf k}\delta(\omega-\tilde E^2_{\mathbf k})]\nonumber\\
&& \quad +\frac{1}{N} \sum_{\mathbf{q}}\Big[|\tilde{u}^+_{\mathbf{kq}}|^2
(1+\tilde{n}^B_{\mathbf{-q}}- \tilde{n}^c_{\mathbf{k+q}})
\delta(\omega-\tilde{\varepsilon}^c_{\mathbf{k+q}}-\tilde{\omega}^{}_{\mathbf{-q}})\nonumber\\
&& \quad +|\tilde{u}^-_{\mathbf{kq}}|^2
(\tilde{n}^B_{\mathbf{-q}}+\tilde{n}^c_{\mathbf{k+q}})
\delta(\omega-\tilde{\varepsilon}^c_{\mathbf{k+q}}+\tilde{\omega}^{}_{\mathbf{q}})\Big]\,.
\end{eqnarray}
The first line in both spectral functions is the coherent contribution which describes 
excitations at the electronic quasiparticle energies $\tilde E_{\bf k}^{(1,2)}$. The remaining lines are 
incoherent contributions. 
They are induced by the electron-phonon 
interaction and turn out to be small.  Note that the coherent part in both cases reduce to 
the mean-field result, when the renormalized quantities are replaced 
by the unrenormalized quantities.  \\

The phonon spectral function $C_{\bf q}( \omega)$ is found in the same way. It is defined  by
 \begin{eqnarray}
\label{A51}
C_{\bf q}(\omega) &=& \frac{1}{2\pi \omega}\int_{-\infty}^\infty \langle [b_{\bf q}(t), b^\dag_{\bf q}] \rangle
 e^{i\omega t} dt  
\end{eqnarray}
where the expectation value is again formed with the full Hamiltonian. 
Using again the unitary invariance, one finds by help of Eqs.~\eqref{A35}, \eqref{A22} 
\begin{align}
\label{A52}
C_{\mathbf q}( \omega)=&\frac{|\tilde{z}_{\mathbf{q}}|^2}{\tilde{\omega}^{}_{\mathbf q}}
\delta(\omega-\tilde{\omega}^{}_{\mathbf q})\\
&+\frac{1}{N}\sum_{\mathbf{k}}\Big[|\tilde{v}^+_{\mathbf{k,-q}}|^2\frac{\tilde{n}^c_{\mathbf{k-q}}-\tilde{n}^f_{\mathbf{k}}}
{\tilde{\varepsilon}^f_{\mathbf{k}}-\tilde{\varepsilon}^c_{\mathbf{k-q}}}
\delta(\omega-\tilde{\varepsilon}^f_{\mathbf{k}}+\tilde{\varepsilon}^c_{\mathbf{k-q}})\nonumber\\
&+|\tilde{v}^-_{\mathbf{kq}}|^2\frac{\tilde{n}^f_{\mathbf{k}}-\tilde{n}^c_{\mathbf{k+q}}}
{\tilde{\varepsilon}^c_{\mathbf{k+q}}-\tilde{\varepsilon}^f_{\mathbf{k}}}
\delta(\omega-\tilde{\varepsilon}^c_{\mathbf{k+q}}+\tilde{\varepsilon}^f_{\mathbf{k}})\Big]\,.\nonumber
\end{align}
The first term on the right-hand side describes a coherent phonon with excitation energy 
at $\omega=\tilde{\omega}^{}_{\mathbf q}$. The remaining terms are
incoherent contributions due to particle-hole excitations of $c$ and  $f$ electrons. 
As before, they are induced by the electron-phonon interaction.
Note that result \eqref{A52} reduces to the mean field result from Sec.~\ref{S:III},
 when the incoherent part  is neglected and the coherent  excitation energy $\tilde \omega_{\bf q}$ is replaced by the unrenormalized energy $\omega_0$ and $\tilde z_{\bf q}$ by 1.  The numerical evaluation of $C_{\bf q}(\omega)$
shows that the weight of the incoherent excitation is small. This allows to use  the energies 
$\tilde{\varepsilon}^c_{\bf k}$ and $\tilde{\varepsilon}^f_{\bf k}$ in expression \eqref{A52}
instead of the correct quasiparticle energies  $\tilde E_{\bf k}^{(1,2)}$.

The numerical solution of the phonon spectral function in Fig.~\ref{fig:Ap} has
shown that $\tilde{\omega}_{\bf q}$ hardens in the non-adiabatic case $(\omega_0=2.5)$ whereas it softens 
in the adiabatic case ($\omega_0=0.5$).
One may ask, can this opposite tendency of the phonon mode already be understood in perturbation theory? 
At first, note that $\omega_0$ is not altered by the mean-field theory, as shown in Sec.~\ref{S:III}. Therefore, the
renormalization of the phonon frequency as well as the $\bf q$-dependence of $\tilde{\omega}_{\bf q}$ 
can only be  caused by the coupling 
to the electronic degrees of freedom, i.e.~by the influence of $\mathcal H_1$. The easiest way to derive 
$\tilde \omega_{\bf q}$ in perturbation theory is to start  from the renomalization equation \eqref{A13} for $\omega_{\bf q, \lambda}$, when the renormalization 
from the original cutoff $\Lambda$ to $\lambda =0$ is done in one single step. Therefore, choosing 
$\lambda = \Lambda$  and also $\Delta \lambda = \Lambda$, one finds from Eq.~\eqref{A13}
\begin{align}
\label{A53}
& \tilde \omega_{\bf q} - \omega_0 = - \frac{2g^2}{N} \sum_{\bf k} \Big( A^+_{\bf k, -\bf q }(\Lambda, \Lambda) (n^f_{\bf k} - n^c_{\bf k -\bf q}) + \nonumber \\
& \hspace*{1.6cm}
+   A^-_{\bf k, \bf q }(\Lambda, \Lambda) (n^f_{\bf k} - n^c_{\bf k +\bf q})
\Big) \nonumber \\
& = - \frac{2g^2}{N} \sum_{\bf k} 
\Big( \frac{n^f_{\bf k} - n^c_{\bf k -\bf q}}{\varepsilon^c_{\bf k - \bf q} - \varepsilon^f_{\bf k} + \omega_0}
+  \frac{n^f_{\bf k} - n^c_{\bf k +\bf q}}{\varepsilon^c_{\bf k + \bf q} - \varepsilon^f_{\bf k} - \omega_0}
\Big) \, , \nonumber \\
& 
\end{align}
which is the perturbative result up to ${\cal O}(g^2)$.
Here we have used that according to Eq.~\eqref{A9} the coefficients 
$A^{\pm}_{\bf k \bf q}(\Lambda, \Lambda)$ reduce to $A^{\pm}_{\bf k \bf q}(\Lambda, \Lambda) = 1/(\varepsilon^c_{\bf k - \bf q} - \varepsilon^f_{\bf k} \pm \omega_0)$.

In the anti-adiabatic case ($ \omega_0 \gg \varepsilon^c_{\bf k - \bf q} , \varepsilon^f_{\bf k}$) Eq.~\eqref{A53}
reduces to 
 \begin{eqnarray}
\label{A54}
\tilde \omega_{\bf q} &=& \omega_0 + \frac{2g^2}{N\omega_0} \sum_{\bf k} \big( n^c_{\bf k -\bf q} +   n^c_{\bf k +\bf q} 
\big) \, .
\end{eqnarray}
Since the second term on the right hand side is positive one indeed finds a hardening of the 
phonon mode. 
In the opposite limit  
 ($ \omega_0 \ll \varepsilon^c_{\bf k - \bf q} , \varepsilon^f_{\bf k}$) the frequency $\omega_0$ can be 
neglected in both denominators of Eq.~\eqref{A53}. Then one arrives at 
 \begin{eqnarray}
\label{A55}
\tilde \omega_{\bf q} &=& \omega_0 - \frac{2g^2}{N} 
\sum_{\bf k} \Big( \frac{n^f_{\bf k }- n^c_{\bf k -\bf q}}{\varepsilon^c_{\bf k -\bf q} - \varepsilon^f_{\bf k}}    +   
 \frac{n^f_{\bf k}- n^c_{\bf k +\bf q}}{\varepsilon^c_{\bf k + \bf q} - \varepsilon^f_{\bf k}} 
\Big) \, . \nonumber \\
&&
\end{eqnarray}
Note that the second term on the right hand side is negative. Thus, the result is a softening of the 
phonon mode in the adiabatic limit, which is like the result in the anti-adiabatic limit in agreement with the 
numerical outcome from the full PRM calculation.

For the numerical evaluation of the various physical quantities from Sec.~\ref{S:V} within the PRM one has to solve 
the sets of renormalization equations \eqref{A11}-\eqref{A15} for the parameters of $\mathcal H_\lambda$ self-consistently
 together with the set  \eqref{A26}-\eqref{A31} for the 
expectation values.   
Starting with some initial values of $n^c_{\mathbf{k}}$, $n^f_{\mathbf{k}}$, $n^B_{\mathbf{-q}}$, and 
$\langle c^\dagger_{\mathbf{k+Q}}f^{}_{\mathbf{k}}\rangle$, the renormalization equations are 
integrated in steps $\Delta\lambda$ little by little until. At $\lambda=0$, the Hamiltonian and all 
quasiparticle operators are completely renormalized and the new expectation values can be calculated.
Then the  renormalization process is restarted. Convergence is assumed to be achieved if all quantities are 
determined with a relative error less than $10^{-5}$. The spectral functions are evaluated with a Gaussian 
energy broadening of 0.06 $t^c=1$. 
In the numerics, we have used a one-dimensional lattice with  $N=1000$ sites. 
$\Delta\lambda$ was customarily chosen as 
$\Delta\lambda \approx 0.01$ (if we use $\Delta\lambda=0.1$ in order to reduce the computational effort the discrepancy is proven to be small, however). 
Concerning the parametrization of the band structure, we  choose $\varepsilon_{}^{f}-\varepsilon_{}^{c}=-1$ (where 
$\varepsilon^c$ was fixed to $\varepsilon^c=0$) 
and $t^{f}=-0.3$ to ensure an (indirect)  semimetallic state for the non-interacting half-filled band case (cf. Fig.~\ref{fig:D-G0}). 
 
\section{Anti-adiabatic limit}
\label{B}
In the anti-adiabatic limit the phonon frequency is assumed to be large compared to the electronic
energies, $\omega_0 \gg \varepsilon^f_{\bf k}, \varepsilon^c_{\bf k}$. 
One realizes that the renormalization due to the elimination of $\mathcal H_1$ becomes rather small  in this limit.  
This follows from expression \eqref{A8} for $X_{\lambda, \Delta \lambda}$ which has 
coefficients 
\begin{eqnarray}
  \label{B1} 
&& A^{\pm}_{\bf k \bf q}(\lambda, \Delta \lambda) =
\frac{\Theta^\pm _{\bf k \bf q, \lambda} \big( 1- \Theta^\pm_{\bf k \bf q, \lambda- \Delta \lambda}\big)}
{\varepsilon^c_{\bf k + \bf q, \lambda} - \varepsilon^f_{\bf k, \lambda} \pm \omega_{\mp\bf q, \lambda}}\, .
 \end{eqnarray}
For large $\omega_{\bf q, \lambda} \sim \mathcal O(\omega_0)$, expression \eqref{B1} 
reduces to 
\begin{eqnarray}
  \label{B2} 
&& A^{\pm}_{\bf k \bf q}(\lambda, \Delta \lambda) \approx
\frac{\Theta^\pm _{\bf k \bf q, \lambda} \big( 1- \Theta^\pm_{\bf k \bf q, \lambda- \Delta \lambda}\big)}
{ \pm \omega_{0}}\, ,
 \end{eqnarray}
where the $\Theta$-functions are now independent of $\bf k$:
\begin{eqnarray}
 && \label{B3} 
\Theta^\pm _{\bf k \bf q, \lambda} \approx \Theta(\lambda - |\omega_0|) =: \Theta_{\bf q,\lambda} \,.
 \end{eqnarray}
Obviously, for large energy $\omega_0$, Eq.~\eqref{B2} only allows small 
renormalization contributions. Moreover, 
the product $ \Theta_{\bf q,\lambda}( 1-  \Theta_{\bf q,\lambda-\Delta \lambda})$ in the numerator
prevents any $\bf k$-dependent renormalization contribution between cutoff $\lambda$ 
and $\lambda- \Delta \lambda$.
 \\

(i) {\it Gap equation}:  For  the conditional equation
of the order parameter $\Delta$, obtained by the PRM, one concludes that  in the anti-adiabatic limit
it reduces to the mean-field expression \eqref{9h} since renormalization contributions  due to $\mathcal H_1$ 
are suppressed. Thus, for $\omega_0 \rightarrow \infty$ one obtains as asymptotic result  
\begin{eqnarray}
\label{B4}
1= \frac{4g^2}{\omega_0} \frac{1}{N} \sum_{\bf k} \textrm{sgn} (\varepsilon^f_{\bf k} - \varepsilon^c_{\bf k + \bf Q})
\frac{f^F(E_{\bf k}^{(1)})- f^F(E_{\bf k}^{(2)})}{W_{\bf k}} \, , \nonumber \\
&&
\end{eqnarray}
where all quantities are unrenormalized. 

At the critical electron-phonon coupling $g= g_c$ the order parameter $\Delta$ vanishes and the condition  
\eqref{B4} becomes
\begin{eqnarray}
\label{B5}
1= \frac{4g_c^2}{\omega_0} \frac{1}{N} \sum_{\bf k} 
\frac{f^F(\varepsilon^f_{\bf k})- f^F(\varepsilon^c_{\bf k + \bf Q})}
{\varepsilon^c_{\bf k + \bf Q} - \varepsilon^f_{\bf k}} \, . \nonumber \\
&&
\end{eqnarray}
 In this extreme anti-adiabatic limit, the squared critical coupling $g^2_c$ scales  linearly with $\omega_0$, provided the electronic parameters in the sum over $\bf k$ are kept constant.  
 This behavior is in perfect agreement with the outcome from the numerical solution of the 
 PRM equations,  as shown in the inset of  Fig.~\ref{fig:Acf1}. Even, the numerical result from \eqref{B4}
  $(g^2_c/\omega_0)_{\omega_0\to\infty} \simeq 0.16$  is in acceptable agreement with that 
from Sec.~\ref{S:V}, which is $(g^2_c/\omega_0)_{\omega_0\to\infty} \simeq 0.14$. 

(ii) {\it Incoherent excitations}: 
The above feature of $ A^{\pm}_{\bf k \bf q}(\lambda, \Delta \lambda)$
can also be used to explain the behavior of the incoherent contributions to  the electronic spectral functions 
$A^{(c,f)}_{\bf k}(\omega)$  in Figs.~\ref{fig:Acf1} and \ref{fig:Acf2}. As long as the electron-phonon coupling $g$ 
is not too small,  the figures show that for the non-adiabatic case 
($\omega_0 = 2.5$) in Fig.~\ref{fig:Acf2} the incoherent contributions to the spectral functions are 
less pronounced than for the adiabatic case ($\omega_0=0.5$) in Fig.~\ref{fig:Acf1}. This property results from the
prefactors $|\tilde t^{\pm}_{\bf k -\bf q, \bf q}|^2$ and  $|\tilde u^{\pm}_{\bf k, \bf q}|^2$ of the incoherent contributions 
in Eqs.~\eqref{A51}and \eqref{A52}. They are found from the solution of the renormalization 
equations \eqref{A26} and \eqref{A27} for $ t^{\pm}_{\bf k -\bf q, \bf q, \lambda}$ and $ u^{\pm}_{\bf k, \bf q, \lambda}$, 
and are governed by the coefficients $A^{\pm}_{\bf k \bf q}(\lambda, \Delta \lambda)$.
A similar behavior is also observed for the phonon spectral function $C_{\bf q}(\omega)$
in Fig.~\ref{fig:Ap}, where the incoherent contributions in the anti-adiabatic regime 
are strongly suppressed. Note that also the weight of the incoherent excitations increases when the coupling 
$g$ is increased. 

\end{appendix}



\bibliography{ref}
\end{document}